\documentclass{aa}  
\usepackage{longtable}
\usepackage{lscape}
\usepackage[usenames, dvipsnames]{color}
\usepackage{booktabs, caption, makecell}
\usepackage{epstopdf}
\usepackage{gensymb}
\usepackage{threeparttable}
\usepackage{geometry}
\usepackage{hyperref}

\graphicspath{ {plots/}  }
\usepackage{epstopdf}

\definecolor{mypink3}{cmyk}{0.26, 1.0, 0.0, 0.56}
\definecolor{mygreen1}{rgb}{0.314, 0.588, 0.078}
\definecolor{gray}{rgb}{0.5, 0.5, 0.5}

\usepackage{color}

\usepackage{graphicx}
\usepackage{txfonts}

\begin{document}

   \title{Dust masses of young disks: constraining the initial solid reservoir for planet formation}

   \author{Łukasz Tychoniec \inst{1}, Carlo F. Manara \inst{2}, Giovanni P. Rosotti\inst{1}, Ewine F. van Dishoeck \inst{1,3}, Alexander J. Cridland \inst{1}, \\ Tien-Hao Hsieh\inst{4},
   Nadia M. Murillo \inst{5}, Dominique Segura-Cox\inst{3}, Sierk E. van Terwisga\inst{6}, John J. Tobin\inst{7}
          }
   \institute{Leiden Observatory, Leiden University, PO Box 9513, NL-2300RA, Leiden, The Netherlands\\
              \email{tychoniec@strw.leidenuniv.nl}
        \and 
        European Southern Observatory, Karl-Schwarzschild-Strasse 2, D-85748 Garching bei M{\"u}nchen, Germany
        \and 
Max-Planck-Institut f{\"u}r Extraterrestrische Physik, Giessenbachstrasse 1, D-85748 Garching bei M{\"u}nchen, Germany
        \and
        Institute of Astronomy and Astrophysics, Academia Sinica, P.O. Box 23-141, Taipei 106, Taiwan
        \and 
        The Institute of Physical and Chemical Research (RIKEN), 2-1, Hirosawa, Wako-shi, Saitama 351-0198, Japan
        \and Max Planck Institute for Astronomy, Königstuhl 17, D-69117 Heidelberg, Germany
        \and National Radio Astronomy Observatory, Charlottesville, VA 22903, US
             }

  \abstract
  {In recent years evidence has been building that planet formation starts early, in the first $\sim$ 0.5 Myr. Studying the dust masses available in young disks enables understanding the origin of planetary systems since mature disks are lacking the solid material necessary to reproduce the observed exoplanetary systems, especially the massive ones.
  }
   { We aim to determine if disks in the embedded stage of star formation contain enough dust to explain the solid content of the most massive exoplanets.}
   {We use Atacama Large Millimeter/submillimeter Array (ALMA) Band 6 (1.1 -- 1.3 mm) continuum observations of embedded disks in the Perseus star-forming region together with Very Large Array (VLA) Ka-band (9 mm) data to provide a robust estimate of dust disk masses from the flux densities measured in the image plane.}
   {A strong linear correlation between the ALMA and VLA fluxes is observed, demonstrating that emission at both wavelengths is dominated by dust emission. For a subsample of optically thin sources, we find a median spectral index of 2.5 from which we derive the dust opacity index $\beta = 0.5$, suggestive of significant dust growth. Comparison with ALMA surveys of Orion shows that the Class I dust disk mass distribution between the two regions is similar, but that the Class 0 disks are more massive in Perseus than those in Orion. Using the DIANA opacity model including large grains, with a dust opacity value of $\kappa_{\rm 9\ mm}$  = 0.28 cm$^{2}$ g$^{-1}$, the median dust masses of the embedded disks in Perseus  are 158 M$_\oplus$ for Class 0 and 52 M$_\oplus$ for Class I from the VLA fluxes. The lower limits on the median masses from ALMA fluxes are 47 M$_\oplus$ and 12 M$_\oplus$ for Class 0 and Class I, respectively, obtained using the maximum dust opacity value $\kappa_{\rm 1.3mm}$  = 2.3 cm$^{2}$ g$^{-1}$. The dust masses of young Class 0 and I disks are larger by at least a factor of 10 and 3, respectively, compared with dust masses inferred for Class II disks in Lupus and other regions.}
   {The dust masses of Class 0 and I disks in Perseus derived from the VLA data are high enough to produce the observed exoplanet systems with efficiencies acceptable by planet formation models: the solid content in observed giant exoplanets can be explained if planet formation starts in Class 0 phase with an efficiency of $\sim$ 15\%. A higher efficiency of $\sim$ 30\% is necessary if the planet formation is set to start in Class I disks.}

   \titlerunning{Dust masses of young disks}
   \authorrunning{Tychoniec et al.}
   \maketitle

\section{Introduction}

The formation of planets is inherently entangled with the formation and evolution of their natal protoplanetary disks. The physical conditions and chemical composition at the onset of planet formation determine the properties of the resulting planetary systems \citep[e.g.,][]{Armitage2011,Oberg2011b, Morbidelli2016}. The key question is then: at what stage of disk evolution do planets start to form?

The protoplanetary disks around Class II pre-main sequence stars were considered to be the starting point of the planet formation process. However, submillimeter surveys of those disks reveal that the mass reservoir available in Class II disks is much lower than the masses needed to explain the formation of the observed exoplanetary systems \citep{Andrews2007,Greaves2010, Williams2012,Najita2014, Manara2018}. Structures observed in the disks \citep[e.g.,][]{Marel2013a,Andrews2018, Long2019} are also evidence that planet formation is already underway in the Class II phase. One of the possible solutions to this conundrum is to move the onset of planet formation to the younger disks surrounding Class 0/I protostars  \citep[< 0.5 Myr;][]{Dunham2014a}, where more material is available \citep{Andrews2007a, Greaves2011, Ansdell2017}. 

There is other evidence for early planet formation. The distribution of different types of meteorites in our solar system can be explained by the formation of Jupiter's core in the first million years of the solar system \citep{Kruijer2014}. There is also evidence for dust growth in the earliest stages of disk formation \citep[e.g.,][]{Jorgensen2007,Kwon2009,Miotello2014,Harsono2018,Hsieh2019a}.
Another indication is provided by young sources with structures suggestive of ongoing planet formation \citep[e.g.,][]{,ALMA2015, Sheehan2018}. These all suggest that planet formation starts early in disks surrounding much younger Class 0 and Class I protostars rather than in Class II disks. 

This begs the question: what is the amount of material available for planet formation in Class 0/I disks? \cite{Greaves2011}, in a study of a small sample of Class 0 disks known at the time, found that 20 - 2000  M$_\oplus$ dust mass is available in Class 0 disks around low-mass stars; they concluded that this is sufficient to form the most massive exoplanet systems found to date. An analysis of a sample of Class I disks in Taurus \citep{Andrews2013} combined with information about the occurrence of exoplanets led \cite{Najita2014} to conclude that Class I disks can explain the population of exoplanetary systems, contrary to Class II disks in the same region. A study of a larger sample of young disks extending to Class 0 protostars is needed to put constraints on planet formation timescales and efficiency.

In the first complete survey of Class 0/I protostars in a single cloud, Perseus, \cite{Tychoniec2018a} used Very Large Array (VLA) 9 mm observations at 75 au resolution to show that there is a declining trend in the dust masses from Class 0 to Class I disks. The median masses for the Class 0 and Class I phase ($\sim$250 and $\sim$100 M$_\oplus$, respectively) are explained by a significant fraction of the dust being converted into larger bodies already in the Class 0 phase. Moreover, they compared the results for Class 0/I disks in Perseus with Atacama Large Millimeter/submillimeter Array (ALMA) observations of several Class II regions which have mean dust masses in the range 5-15 M$_\oplus$ \citep{Ansdell2017}. This suggests that dust masses in the Class 0/I disks are an order of magnitude higher that those for Class II disks. Note, however, that the adopted dust mass absorption coefficient ($\kappa_{\nu}$ - dust opacity) varies in these studies.

ALMA observations by \citet{Tobin2020} in the Orion Molecular Cloud, based on the largest sample of protostars observed in a single region at sub-millimeter wavelengths (379 detections), found much lower mean dust masses for Class 0 and I disks than those in Perseus, 26 and 15 M$_\oplus$, respectively. Very low Class I mean dust disk mass (4 M$_\oplus$) were also reported in the Ophiuchus star-forming region
\citep{Williams2019}. Also in this case, different opacities assumed in those studies could contribute to the difference between the median masses measured.

Comparison of the VLA observations for Perseus
with other embedded disks surveys using ALMA is difficult because of
the different wavelength range of observations. The VLA obsercvations
at 9 mm can have a significant free-free emission contribution, which
could result in overestimating the actual flux coming from the dust
\citep[e.g.,][]{Choi2009}; although \cite{Tychoniec2018a} applied the correction for a free-free contribution using information from the C-band (4.1 and 6.4 cm) flux densities. On the other hand, the dust emission at those long
wavelengths is less likely to be optically thick than that
in the ALMA wavelength range \citep{Dunham2014b}. The way forward is to use observations of
young disks with VLA and ALMA in the same star-forming region, offering
a direct comparison of dust disk masses and determining if the difference in observing wavelengths can be the reason for the described
differences. Therefore, in this work we present ALMA observations of protostars in Perseus and compare them with our previous VLA data.

This work aims to compare the solid masses of the embedded (Class 0/I) disks with the masses of the exoplanetary systems observed to date to ultimately infer an efficiency of the planet formation. In Section 2, we describe the ALMA observations and data analysis. In Section 3, the integrated fluxes at 1 mm and 9 mm are compared, and dust masses are calculated based on those fluxes and then compared with other young and more mature dust disks observed with ALMA. In Section 4, we put the inferred masses in the context of known exoplanetary systems masses and planet formation models.

\section{Observations and analysis}
\subsection{Observations}
In this paper we analyze ALMA Band 6 continuum observations of 44 protostars in the Perseus molecular cloud. The data were obtained in September 2018 with a Cycle 5 program (2017.1.01693.S, PI: T. Hsieh). The absolute flux and bandpass calibrator was J0237+2848, and the phase calibrator was J0336+3218. Continuum images and spectral lines observed in this project are presented in \cite{Hsieh2019}. The continuum bandwidth was $\sim 1.85\,$GHz centered at 267.99 GHz (1.1 mm). The absolute flux calibration uncertainty is on the order of $\sim$ 30\%. The synthesized beam of the continuum observations in natural weighting is $0\farcs45 \times 0\farcs30$. The average spatial resolution of observations ($0\farcs38$)  corresponds to 110 au (diameter) at the distance to Perseus (293$\pm 22$ pc; \citealt{Ortiz-Leon2018}). The typical rms value of the continuum images is  $\sim$ 0.1 mJy beam$^{-1}$.
 
 Additional data on 8 disks were obtained in a Cycle 5 program (2017.1.01078.S, PI: D. Segura-Cox). The continuum bandwidth was centered at 233.51 GHz (1.3 mm) with a total bandwidth of 2 GHz. The average synthesized beam of 0\farcs41$\times$0\farcs28 provides spatial resolution  corresponding to 100 au at the distance of Perseus. The rms value of the images is $\sim$ 0.05 mJy beam$^{-1}$. The absolute flux and bandpass calibrator was J0510+1800 and the phase calibrator was J0336+3218. The accuracy of the flux calibration is on the order of $\sim$ 10\%. The measurement sets were self-calibrated and cleaned with the robust parameter 0.5.

We also use the flux densities of 25 disks published in \cite{Tobin2018} which were observed at 1.3 mm with a resolution of $0\farcs27\times0\farcs16$ and sensitivity of 0.14 mJy beam$^{-1}$.  The flux and disk masses in \cite{Tobin2018} are measured using a Gaussian fit in the image domain to the compact component in the system without subtraction of an envelope component. Altogether we compile a sample of 77 Class 0 and Class I disks in Perseus observed with ALMA.  In the following, when referring to ALMA data, we use 1 mm observations for short, but anywhere the wavelength is used to calculate properties of the source (e.g., disk mass) the exact value of the observed wavelength is used.

 The VLA observations come from the VLA Nascent Disks and Multiplicity Survey (VANDAM) \citep{Tobin2015a, Tobin2016, Tychoniec2018}. The sample for the VANDAM survey was prepared based on unbiased infrared and submillimeter surveys of protostars in Perseus \citep{Enoch2009,Evans2009, Sadavoy2014}. Fluxes at 9.1 mm (Ka-band), obtained with 0.25\arcsec resolution from 100 Class 0 and I disks (including upper limits) were reported in \cite{Tobin2016}. \cite{Tychoniec2018} applied a correction for free-free emission, based on C-band (4.1 and 6.4 cm) observations. In that work, all sources with a Ka-band spectral index suggestive of emission not coming from dust ($\alpha \ll 2$) were marked as upper limits, and we use the same criteria here. We use the 9 mm fluxes corrected for the free-free emission for further analysis.

\subsection{Gaussian fitting}

Pre-ALMA surveys of embedded sources have found that disk masses are
typically only a small fraction of the total envelope mass in the
Class 0 phase (1-10\%), becoming more prominent as the system evolves
in the Class I phase (up to 60\%, e.g., \citealt{Jorgensen2009}). In the
much smaller ALMA beam, the envelope contamination is reduced
\citep[e.g.,][]{Crapsi2008}, but still needs to be corrected for \citep{Tobin2020}. Here both components, disk and envelope, are represented by Gaussians.

The CASA \citep{McMullin2007} v. 5.4.0 $\it imfit$ task was used to fit Gaussian profiles to the sources. After providing the initial guess, all parameters: position, flux, and shape of the Gaussian, were set free during the fit.  All sources were inspected by eye to assess the number of necessary Gaussian components. In case of a single source without a noticeable contribution from the envelope, a single compact Gaussian with the size of the synthesized beam  was provided as input to the $\it imfit$ task (Fig. \ref{fig:fit_examples}a). In cases where a contribution of the envelope by eye was significant, an additional broad Gaussian with a size of  3\arcsec was added to the initial guess parameters of the fitting (Fig. \ref{fig:fit_examples}b). 
In two cases (Per-emb-4 and SVS13A2) it was necessary to fix the size of the Gaussian to the synthesized beam size for the fit to converge (Fig. \ref{fig:fit_examples}c). Two binary systems with separations below our resolution (Per-emb-2 and Per-emb-5) are treated as single systems with a common disk.

The flux density of the compact Gaussian is assumed to be that of the embedded disk. It is called `disk' here, even though no evidence for a Keplerian rotation pattern yet exists. We report the measured fluxes of the embedded disks in Table \ref{tab:table1}.

The broad component is used only to force the $\it imfit$ task to not fit extended emission without constraining the compact Gaussian size which would in turn overestimate the flux of the compact emission. It was necessary to add an envelope component to 31 sources out of 51 targets, specifically 20 Class 0 and 11 Class I sources. We assessed remaining 6 Class 0 sources and 20 Class I sources as not having significant contribution from their envelope. 

The envelope flux remaining after subtracting the model of the disk component is measured as the flux in the area of the size of the FWHM of the disk in the residual image. This ratio of the envelope residual flux to the disk flux ranges from less than 1\% to usually below 30\%. In one case the source is dominated by the envelope emission (Per-emb-51; Fig. \ref{fig:fit_examples}d), but after the envelope component subtraction the residual flux is only $\sim\ 6\%$ of that of the disk (Table \ref{tab:table1}).  The two sources with high values of the ratio - Per-emb-22-B and Per-emb-27B are heavily affected by the nearby binary component so the value is not reliable.

We stress that the remaining envelope fraction is not incorporated in the flux density of the disk component, and it is presented to show that fitting the envelope component is needed to exclude the contamination of the envelope from the disk. Fig. \ref{fig:fit_examples} shows that residuals are significantly reduced after removing the envelope contribution, and that without fitting the extended component some of this emission could contaminate the flux coming from the disk.

\section{Dust disk masses}
\subsection{Comparison of the integrated fluxes between 1 and 9 mm }

Measurements of the continuum emission at different wavelengths allow us to analyze the properties of the emitting material. First, it is important to verify that the fluxes at both 1 mm (ALMA) and 9 mm (VLA) have their origin in the same physical process. This is to confirm that the correction for contamination of the VLA observations by the free-free emission is accurate. In order to do so, we investigate the correlation between the flux densities at both wavelengths and the spectral index of the emission for each source.

\begin{figure}[h]
\centering
  \includegraphics[width=0.95\linewidth,trim={0cm 0cm 0cm 2cm},clip]{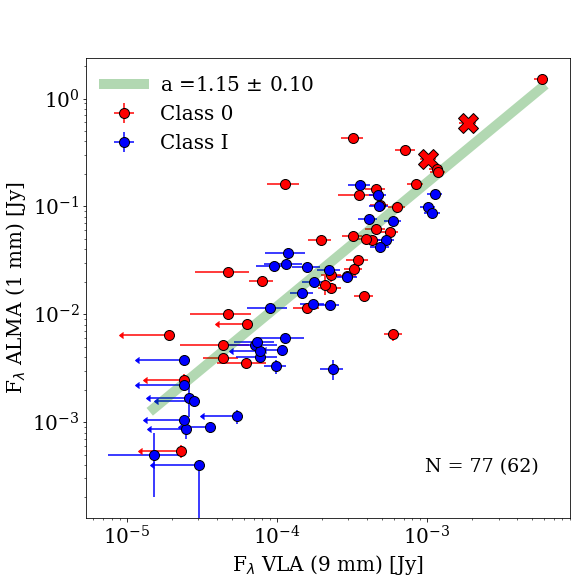}
  \caption{ALMA (1.1 and 1.3 mm) integrated fluxes plotted against VLA integrated fluxes at 9.1 mm. A total of 77 sources are plotted but only the 62 sources that are not upper limits are included in fitting the linear function. The best fit to the data is shown with the green line and has a slope of 1.15 $\pm$ 0.1. The Class 0 sources are shown in red and Class I sources in blue. Crosses mark sources that are unresolved binaries. }
 \label{fig:fluxes_ALMA_VLA}
\end{figure}

The flux densities from the ALMA 1.1-1.3 mm observations are presented in Table \ref{tab:table1}. In Fig. \ref{fig:fluxes_ALMA_VLA}  we compare the measurements with the VLA 9.1 mm observations \citep{Tobin2016, Tychoniec2018a}. There is a clear correlation with a close-to-linear slope (1.15 $\pm$ 0.10) obtained with the {\it lmfit} Python function.  The fitting was performed excluding upper limits. The value of the slope indicates that all sources have a similar spectral index between ALMA and VLA wavelengths. Thus, the mechanism responsible for emission at both wavelength ranges is the same. Since it is generally accepted that the ALMA 1 mm emission is dominated by dust thermal emission, we can conclude that this is also the case for the 9 mm VLA observations.  The sources with resolved emission at 9 mm can be modelled successfully with the disk \citep{Segura-Cox2016}.

The nature of this emission can be also investigated with the value of the spectral index between the two wavelengths. In the Rayleigh-Jeans approximation, the flux density $F_\nu$ changes with frequency $F_\nu \sim \nu^{\alpha}$, where  $\alpha$ is the spectral index. The dust emissivity index $\beta$ defines the dependence of the dust opacity on the frequency $\kappa_{\nu} \sim \nu^{\beta}$. From the observed spectral index, the emissivity index can be derived accordingly:
\begin{equation}
\label{eq:beta}
\beta = (\alpha - 2)(1+\Delta)\,
\end{equation}
where $\Delta$ is the ratio of optically thick to optically thin emission \citep{Beckwith1990,Lommen2007}. It is often assumed that emission at millimeter wavelengths is optically thin, in which case $\Delta$ = 0 and then $\beta = \alpha - 2$.

The spectral index between the ALMA and VLA fluxes is calculated as

\begin{equation}
\label{eq:spectral_index}
\alpha_{\rm VLA/ALMA}=\frac{\ln (F_{\rm 1\ mm}/ F_{\rm 9\ mm})}{\rm ln(\rm 9\ mm/1\  mm)}.
\end{equation}

The mean spectral index obtained between ALMA and VLA  is $\alpha \sim 2.4$, with a standard deviation for the sample of 0.5. This indicates that $\beta$ = 0.4 $\pm$ 0.5, which is lower than the typical ISM value, i.e. $\sim $1.8 for small grains \citep{Draine2006}. The index is also lower than 1, the value typically used for protoplanetary disks, specifically in our previous study of embedded disks in Perseus \citep{Tychoniec2018a}. If emission is optically thin, the low value  of $\beta$ can point  to dust growth as is commonly seen in Class II disks \citep[e.g.,][]{Natta2004a, Ricci2010, Testi2014}.  There are other effects that could alter the value of the dust spectral index such as dust porosity \citep{Kataoka2014}, and  grain composition \citep{Demyk2017b,Demyk2017a} but to explain  $\beta < 1$ some grain growth is required \citep{Ysard2019}.

While the VLA 9 mm flux is unlikely to be optically thick, the ALMA 1 mm emission from young disks can be opaque. Optically thick emission at 1 mm would result in a lower spectral index value. If the indices obtained  between 1 and 9 mm $\alpha_{\rm VLA/ALMA}$ are consistent with the Ka-band intraband indices $\alpha_{\rm VLA}$, it can be assumed that the emission at 1 mm is optically thin so that it is possible to measure the spectral index in a robust way. We calculate  the VLA intraband spectral index, determined between the two sidebands of the Ka-band observations as follows:

\begin{equation}
\label{eq:spectral_index}
\alpha_{\rm VLA}=\frac{\ln (F_{\rm 8\ mm}/ F_{\rm 10\ mm})}{\rm ln(\rm 10\ mm/8\  mm)}.
\end{equation}

 Fig. \ref{fig:robust_index} shows the range of the $\alpha_{\rm VLA/ALMA}$ and $\alpha_{\rm VLA}$ values measured.
For 23 sources out of 77 we found the  $\alpha_{\rm VLA/ALMA}$ - $\alpha_{\rm VLA}$
$\leq$ 0.4, and therefore in reasonable agreement (see Fig. \ref{fig:SED}). For those sources, the emission at both 1 mm and 9 mm wavelengths is most likely optically thin, so the spectral index should provide information about the grain size. The mean spectral index of those sources is 2.5, which means that $\beta \sim 0.5$. This value confirms that significant dust growth is occurring in the observed disks. The spectral index calculated for the selected optically thin sample (0.5) is similar to that calculated for the full sample (0.4). We therefore proceed with assuming a value of $\beta=$ 0.5 for the further analysis, as an average value, which does not exclude that the 1 mm emission is optically thick. It is also likely that the spectral index varies with the radius of the disk due to optically thick emission close to the protostar and due to the grain growth further out \citep{Pinilla2012,vanTerwisga2018}, as well as a grain size that depends on radius \citep{Tazzari2016}. Our observed emission is largely unresolved and the measured spectral index is an average of those effects.

The emission at shorter wavelengths is more likely to be optically thick. With at least marginally resolved disks, we can obtain an estimate of the dust optical depth, because the extent of the emission allows to approximate the disk radius. We use the major axis deconvolved from the beam as the diameter of the disk. Then, we obtain optical depth as $\tau \sim \kappa_{\nu}\Sigma$, where $\kappa_{\nu}$ is dust opacity used to calculate the mass and $\Sigma$ is averaged surface density. Fig. \ref{fig:dustopacities} presents a distribution of calculated optical depths. For all the disks with available major axis value, we get $\tau < 0.4$ and in vast majortiy of cases $<0.1$. 

Summarizing, we have identified dust thermal emission as the dominating physical process responsible for the emission at both 1 mm and 9 mm. What is more, from the sample of sources for which the emission is most likely optically thin, we calculate a spectral index value of 2.5, suggestive of significant grain growth already at these young stages. 

\subsection{Disk mass measurements }

The continuum flux at millimeter wavelengths is commonly used as a proxy of the dust mass of the emitting region. Here we utilize the collected fluxes for the continuum flux in Perseus with VLA and ALMA to calculate the masses of the embedded disks. The key assumptions used in the calculation, temperature and dust opacity (dust mass absorption coefficient), are discussed. Then we proceed to compare the results with other disk surveys both at Class 0/I and at Class II phases.

From the integrated disk fluxes, the dust mass of the disk is calculated following the equation from \cite{Hildebrand1983}:
\begin{equation}
\label{eq:dustmass}
M=\frac{D^2F_\nu}{{\kappa}_\nu B_\nu(T_{\rm dust})}\,
\end{equation}
where $D$ is the distance to the source, $B_\nu$ is the Planck function for a  temperature $T_{\rm dust}$ and $\kappa_\nu$ is the dust opacity with the assumption of optically thin emission. Temperature of the dust is set to 30 K, typical for dust in dense protostellar envelopes \citep{Whitney2003}, and disks are assumed to be isothermal. The same temperature is set for Class 0 and Class I disks. If the decrease of the temperature of dust from Class 0 to Class I is significant, the mass difference diminishes  \citep[e.g.,][]{Andersen2019}. We consider two cases for the values of $\kappa_\nu$ at 1.3 and 9 mm. 

\begin{figure}[h]
\centering
    \includegraphics[width=0.95\linewidth,trim={0cm 0cm 0cm 0cm},clip]{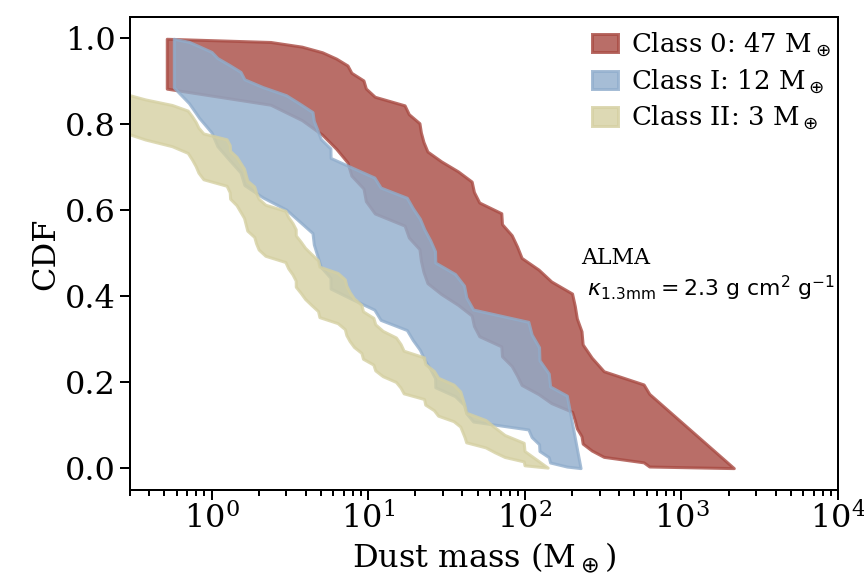}
        \includegraphics[width=0.95\linewidth,trim={0cm 0cm 0cm 0cm},clip]{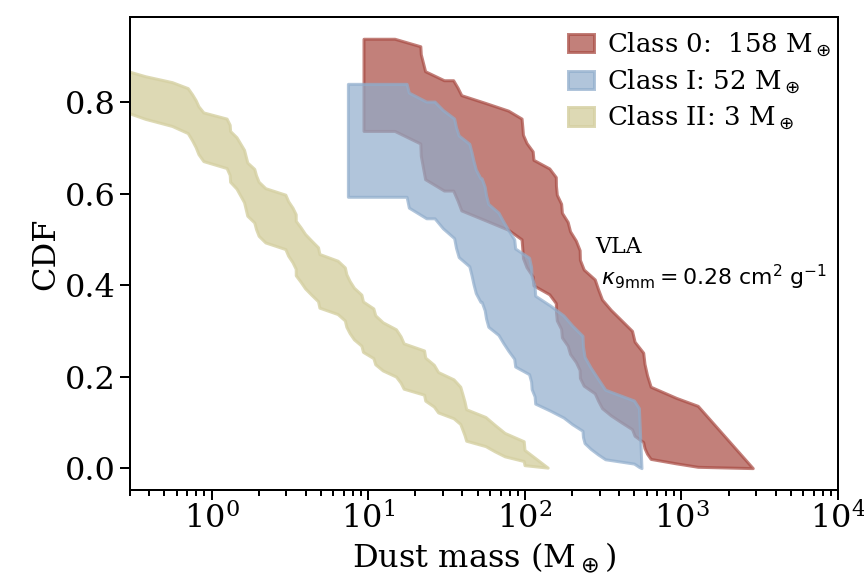}
  \caption{ Cumulative distribution plots of the dust disk masses. {\it Top:} Masses of the Perseus Class 0 and I disks measured with ALMA at 1 mm compared with the Lupus Class II disks measured with ALMA \citep{Ansdell2016}.  The opacity value of $\kappa_{1.3 {\rm mm}}=2.3\ {\rm \ cm}^{2}\ {
  \rm g^{-1}}$ is used to calculate the masses. The ALMA sample consist of 77 sources (38 Class 0 and 39 Class I) and the Lupus sample consist of 69 sources. {\it Bottom:} Masses of the Perseus Class 0 and I disks measured with VLA at 9 mm (red and blue, respectively), compared with the Lupus Class II disks measured with ALMA \citep{Ansdell2016}.  The opacity value of $\kappa_{9 {\rm mm}}=0.28\ {\rm \ cm}^{2}\ {\rm g}^{-1}$ is used to calculate the VLA masses. The VLA sample consist of 100 sources (49 Class 0 and 51 Class I). Medians are indicated in the labels.}
 \label{fig:ALMA_Per_Lup}
\end{figure}

First, since our aim is to compare the results with the Class II disk masses in the literature, most notably the Lupus star-forming region, we use $\kappa_{\rm 1.3\ mm} = 2.3 $~cm$^{2}$~g${^{-1}}$ as used in the determination of masses in \cite{Ansdell2016}.  Fig. \ref{fig:ALMA_Per_Lup} (top panel) shows the cumulative distribution function (CDF) for Class II disks in Lupus and Class 0 and I disks in Perseus, all observed with ALMA. The CDF plot is prepared using the survival analysis with the {\it lifelines} package for Python \citep{DavidsonPilon2017}. The CDF plot describes the probability of finding the element of the sample above a certain value. Uncertainty of the cumulative distribution is inversely proportional to the size of the sample and 1$\sigma$ of the confidence interval is indicated as a vertical spread on the CDF plot. It takes into account the upper limits of the measurement, and the median is only reliable if the sample is complete. While the VLA observations sample is complete, the ALMA sample of disks is not, as we assemble $\sim 80\%$ of the total sample. Therefore the VLA median values and distributions are more reliable.

The median dust mass for young disks in Perseus measured with ALMA at 1 mm with the adopted opacity of $\kappa_{\rm 1.3\ mm} = 2.3 $~cm$^{2}$~g${^{-1}}$ is 47 M${_\oplus}$ and 12 M${_\oplus}$ for Class 0 and Class I disks, respectively (Fig. \ref{fig:ALMA_Per_Lup}, top panel). The median is taken from the value corresponding to the 0.5 probability on the CDF plot. The opacity value used here is likely close to the maximum value of the opacity at 1.3 mm \citep{Draine2006}. In an analysis of dust opacity value at 1.3 mm, \citet{Panic2008} find a range between 0.1 and 2 cm$^{2}$ g$^{-1}$. Therefore, the dust masses obtained with $\kappa_{\rm 1.3\ mm} = 2.3 $~cm$^{2}$~g${^{-1}}$ from \cite{Ansdell2016} should be considered as a lower limit to the disk masses in Perseus. Only if the grain composition is significantly different from the typical assumption, in particular if dust has a significant fraction of amorphous carbon, will the actual masses of the dust be lower by a factor of a few \citep{Birnstiel2018}, even when compared with the $\kappa_{\rm 1.3\ mm} = 2.3 $~cm$^{2}$~g${^{-1}}$ that we assume to provide a lower limit on the dust disk mass.

Regardless of the uncertainties, there is a clear evolutionary trend from Class 0 to Class II with disk masses decreasing with evolutionary phase. The median dust masses for disks in Perseus measured with ALMA of 47 $M{_\oplus}$ and 12 M${_\oplus}$ for Class 0 and Class I disks, respectively, are significantly higher than for Class II disks in Lupus which have a median mass of 3 M${_\oplus}$. We note that this value differs from the 15 M$_\oplus$ value reported in \cite{Ansdell2016}, since in that work the standard mean is calculated, contrary to the median taken from the CDF plot. Also,  distances to Lupus disks have been updated with {\it Gaia} DR2 distances \citep{Gaia2018}. It should be noted that the dust temperature used in \cite{Ansdell2016} was 20 K, while we use 30 K, but the opacity value adopted to calculate the masses is the same. A lower temperature results in an increase of the total mass, based on Equation 4. Therefore if the temperature would be set to 20 K for the Perseus disks, the difference between Class 0/I and Class II disks would be even higher. Class 0/I disks are, however, expected to be warmer than Class II disks \citep{Harsono2015,vantHoff2020}.

As an alternative method, considering that ALMA fluxes can be optically thick, we use the VLA flux densities to estimate the disk masses. Here we adopt $\kappa_{9 {\rm mm}}= 0.28 \ {\rm cm^{2}\ g^{-1}}$ as provided by dust models of the DIANA project \citep{Woitke2016} that consider large grains up to 1 cm; we recall that significant grain growth is indicated in our data by the empirically measured value of $\beta = 0.5$. In \cite{Tychoniec2018a}, a value of $\kappa_{9 {\rm mm}}= 0.13 \ {\rm cm^{2}\ g^{-1}}$ was used, scaled from $\kappa_{1.3 {\rm mm}}= 0.9 \ {\rm cm^{2}\ g^{-1}}$  of \cite{Ossenkopf1994} using $\beta$ = 1. If $\beta$ = 0.5 is instead used to scale the opacity $\kappa_{1.3 {\rm mm}}$ to 9 mm, the value is consistent with that of DIANA.

The median masses measured from the  9 mm observations are 158 M${_\oplus}$ for Class 0 and 52 M${_\oplus}$ for Class I (Fig. \ref{fig:ALMA_Per_Lup}, bottom panel). Those masses are lower than the estimate provided in \cite{Tychoniec2018a} by a factor of two, which stems from the different opacity values used. Additionally the values quoted in \cite{Tychoniec2018a} are regular medians, taken from the sample of detected disks and the distance to the Perseus star-forming region has been revised from 235 to 293 pc \citep{Ortiz-Leon2018} which increases the estimate of the mass.

An important difference between the ALMA and VLA samples is that the VLA sample is complete, as it targeted all known protostars in Perseus \citep{Tobin2016}. Additionally, it is likely that the VLA flux densities are coming from optically thin emission, whereas the ALMA flux densities can become optically thick in the inner regions. We also use a refined model of the dust opacity of the DIANA project \citep{Woitke2016} including large grains. Therefore, the median masses reported with VLA (158 M${_\oplus}$ and 52 M${_\oplus}$ for Class 0 and Class I, respectively) can be considered more robust.

\subsection{ALMA Class 0/I disk masses for different star-forming regions}

Recent ALMA observations of Orion and Ophiuchus reveal masses of embedded Class 0/I dust disks that are somewhat lower than those obtained for Perseus with the VLA \citep{Williams2019, Tobin2020}. Here we collect available ALMA observations for Perseus that use the same techniques as other embedded surveys. Such analysis can reveal the inherent differences between the different protostellar regions.

\begin{figure}[h]
\centering
  \includegraphics[width=0.95\linewidth,trim={0 0cm 0 0cm},clip]{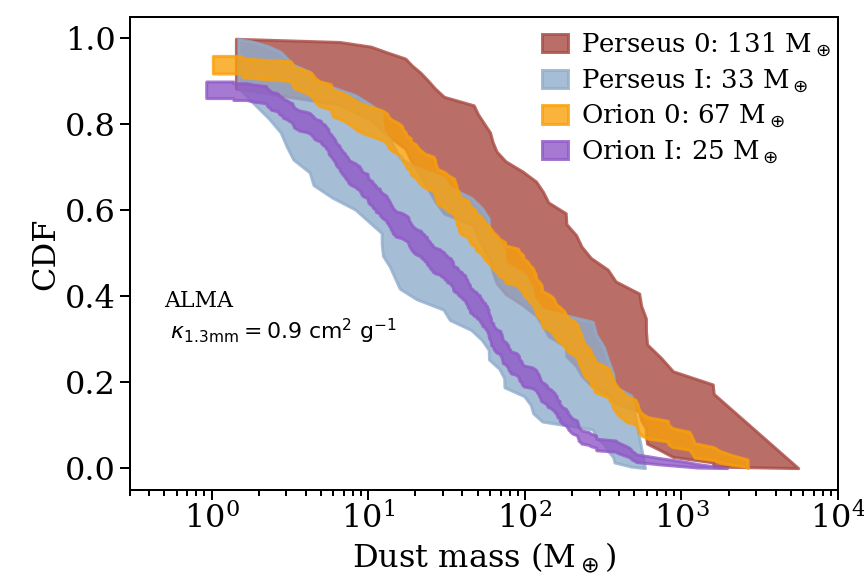}
  \caption{Cumulative distribution plots for Perseus disk masses calculated from ALMA flux densities, and Orion disk masses from \citet{Tobin2020}, both calculated with the same opacity assumptions of $\kappa_{1.3 {\rm mm}}=0.9\ {\rm cm^{2}\ g^{-1}}$. The Perseus sample consist of 77 sources (38 Class 0 and 39 Class I), the Orion sample consist of 415 sources (133 Class 0 and 282). Medians are indicated in the labels.}
 \label{fig:cdplots_per_ori}
\end{figure}

In Fig. \ref{fig:cdplots_per_ori} we show the cumulative distribution of disks observed with ALMA for Orion \citep{Tobin2020} and for Perseus. The Orion disks were targeted within the VANDAM survey of Orion protostars where 328 protostars were observed with ALMA Band 7 (0.87 mm) at 0\farcs1 (40 au) resolution. The sample in \cite{Tobin2020} is divided into Class 0, Class I and Flat spectrum sources. We incorporate the Flat Spectrum sources into Class I in the comparison. An opacity value of $\kappa_{0.87 {\rm mm}}=1.84\ {\rm cm^{2}\ g}^{-1}$  \citep{Ossenkopf1994} has been assumed to calculate the masses in the Orion survey. We use the same reference for opacity at 1.3 mm wavelengths, $\kappa_{1.3 {\rm mm}}=0.9\ {\rm cm^{2}\ g}^{-1}$, to calculate the masses from ALMA flux densities in Perseus to compare with the Orion sample. The median for Class 0 dust disk masses is still significantly lower in Orion, 67 M$_\oplus$ versus 131 M$_\oplus$ in Perseus, but is remarkably similar for Class I: 25 M$_\oplus$ versus 33 M$_\oplus$. Thus, using ALMA-measured flux densities and the same opacity assumption as \citet{Tobin2020}, we find that there are some inherent differences between the population of Class 0 disks in Perseus and Orion. Note that values calculated here are different than reported in \cite{Tobin2020} because the temperature of the dust was scaled with the luminosity in that work, while we use a constant $T$ = 30 K for a consistent comparison with our sample.

Differences in sound speed or initial core rotation can result in different disk masses \citep[e.g.,][]{Terebey1984,Visser2009}. The initial composition of grains could also affect the dust spectral index. The Orion Molecular Complex seems to show a higher fraction of the amorphous pyroxene than the typical ISM \citep{Poteet2012}. It is likely that such factors are resulting in different observed masses between Orion and Perseus. \cite{Tobin2020} noted that the 9 mm flux density distribution is similar between Perseus and Orion.

The low Class I median masses reported in Ophiuchus \citep[][median mass 3.8 M$_\oplus$]{Williams2019} are puzzling as it suggests that the problem of missing dust mass for planet formation extends from Class II to Class I disks. In our data, the median Class I disk mass median is 11 M$_\oplus$ for the same opacity assumption as in \cite{Williams2019}, a factor of 3 higher. The Ophiuchus sample does not include the entire population of Class I disks in Ophiuchus and may be contaminated with more evolved sources due to the high foreground extinction \citep{vanKempen2009a, McClure2010}. For this reason, we will not include it in the further analysis. Despite those caveats it is possible that the population of young disks in Ophiuchus is less massive than in Perseus and Orion.

\section{Exoplanetary systems and young disks - a comparison of their solid content}

Surveys of protoplanetary (Class II) disks around pre-main sequence stars reveal that dust masses of most disks are not sufficient to explain the inferred solid masses of exoplanetary systems \citep{Williams2012, Najita2014, Ansdell2017, Manara2018}. On the other hand, the results from the younger (Class 0/I) star-forming regions show that the dust reservoir available in younger disks is much higher than in the Class II phase \citep{Tychoniec2018a, Tobin2020}. Here, we aim to determine if the amount of dust available at the onset of planet formation (Class 0/I disks) agrees with the masses of the exoplanet systems observed for reasonable efficiencies of the planet formation process. Simply stated: are the masses of the embedded disks high enough to produce the observed population of the most massive exoplanet systems, or does the problem of the missing mass extend even to the youngest disks? 

In this analysis we focus on the Perseus sample, the only complete sample that is available for Class 0/I protostars in a low-mass star-forming region. As such it guarantees that there is no bias towards the more massive disks. Perseus, however, may not be a representative star-forming region for the environment of our own Solar System \citep{Adams2010}. Also, because it is difficult to estimate the stellar mass of the Class 0/I sources, making a comparison of planets and disks around similar stellar types is challenging. Therefore, we include as well a comparison with the Orion disks. The Orion star-forming region contains more luminous protostars than Perseus; thus it might be more representative of the initial mass function. The other limitation is that we analyze mostly unresolved disks, hence the radial dust distribution in the disk is unknown.

\subsection{Exoplanet sample selection}

The exoplanet systems masses were obtained from the exoplanet.eu database \citep{Schneider2011}. From the catalog (updated 28.04.2020) we obtained 2074 exoplanets with provided value for the total mass, either a true mass, or a lower limit to the mass (M$\ \times\ sin(i)$). We do not filter for detection method, mass measurement method, or stellar type of the host star. The mass estimation method for the majority of planets with information on a mass detection method provided is a radial velocity method, which introduces a strong bias toward more massive exoplanets. 
Indeed 1373 of the exoplanets in our analysis have a total (gas+dust) masses above 0.3 Jupiter masses (M$\rm_J$). There are 1062 systems with more than one planet where at least one is > 0.3 M$\rm_J$, and 173 systems with a single > 0.3 M$\rm_J$ planet. 

Gaseous planets are expected to be less frequent than the rocky low-mass planets \citep[e.g.,][]{Mayor2011}. It is estimated that only 17-19\% of planetary systems would contain a planet more massive than  0.3 M$\rm_J$ within 20 au orbit \citep{Cumming2008}. Therefore, we use only systems containing at least one planet with a total mass of 0.3 M$\rm_J$ and normalize it to 18\% of the total population.  This is done by setting the 18\% value of the CDF plot at the estimated solid fraction of the gaseous planet of 0.3 M$_{\rm J}$, which is 27.8 M$_{\oplus}$. We assume that 82\% of the systems have masses below that value. By doing so, we focus on the sample of gas giants and their solid material content. 

In this work, we focus on a reliably estimated solid mass and its cumulative distribution for the most massive exoplanetary systems. By comparing their CDF to the total dust mass distribution from surveys of young disks we can answer the pivotal question: do the Class 0/I disks contain enough solids to explain the masses of those systems.

Our study focuses on the dust masses of disks. In order to compare the solid content between the disks and exoplanets, we calculate the solid content in exoplanets using the formula from  \cite{Thorngren2016} for estimating the solid content in gaseous planets. This study is based on structural and thermal planetary evolution models relating the metallicity of a gas giant with the total mass of the planet. Importantly, the metals in those gas giants are assumed to be located not only in the core but also in the envelope of a planet. We combined the masses of  planets orbiting the same star to retrieve the total dust mass of the system, resulting in 1235 systems in the analysis.

         \begin{figure}[h]
\centering
  \includegraphics[width=0.95\linewidth,trim={0cm 0cm 0cm 0cm},clip]{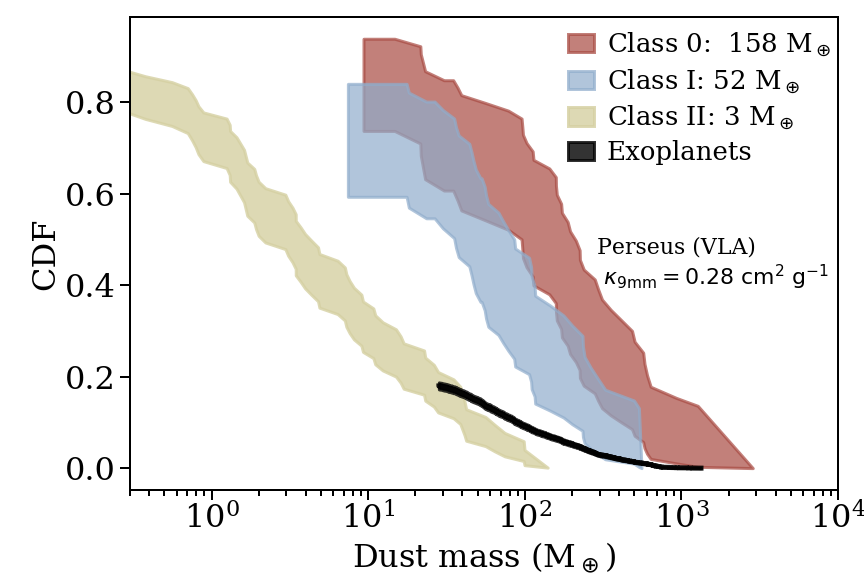}
  
      \includegraphics[width=0.95\linewidth,trim={0cm 0cm 0cm 0cm},clip]{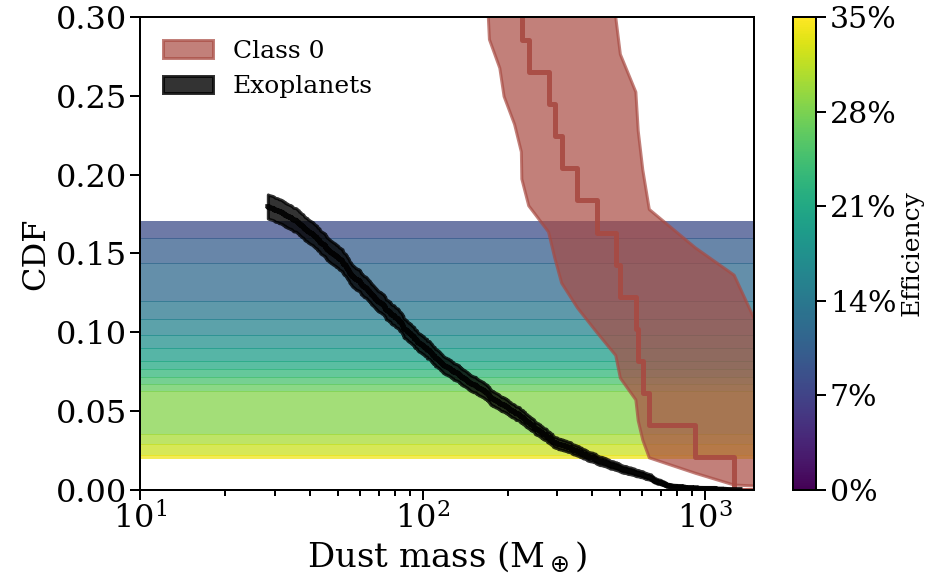}

\caption{Cumulative distribution function of dust disk masses and solid content of exoplanets.
{\it Top:} Cumulative distribution function of dust masses for Class 0 (red) and Class I (blue) disks in Perseus and Class II disks (yellow) in Lupus measured with ALMA \citep{Ansdell2016}. In black, the masses of the exoplanet systems are normalized to the fraction of the gaseous planets \citep{Cumming2008}. Perseus disk masses calculated with  $\kappa_{9 {\rm mm}}=0.28\ {\rm cm^{2}\ g^{-1}}$ from the VLA fluxes. Medians are indicated in the labels. {\it Bottom:} Zoom-in to the ranges where exoplanets are present. The color scale shows the efficiency needed for the planet formation for a given bin of the distribution.}

 \label{fig:hist_exo+disks}
\end{figure}

 \begin{figure}[h]
\centering      
      
    \includegraphics[width=0.95\linewidth,trim={0cm 0cm 0cm 0cm},clip]{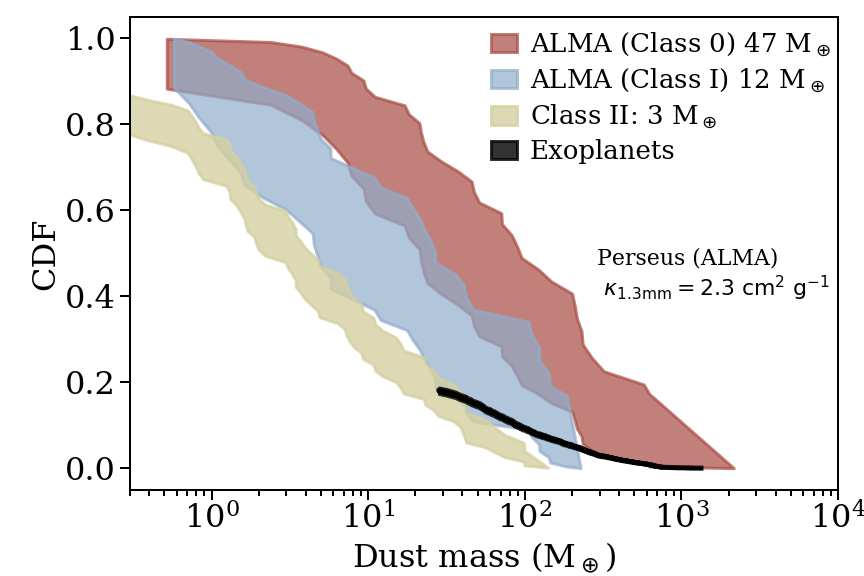}
        \includegraphics[width=0.95\linewidth,trim={0cm 0cm 0cm 0cm},clip]{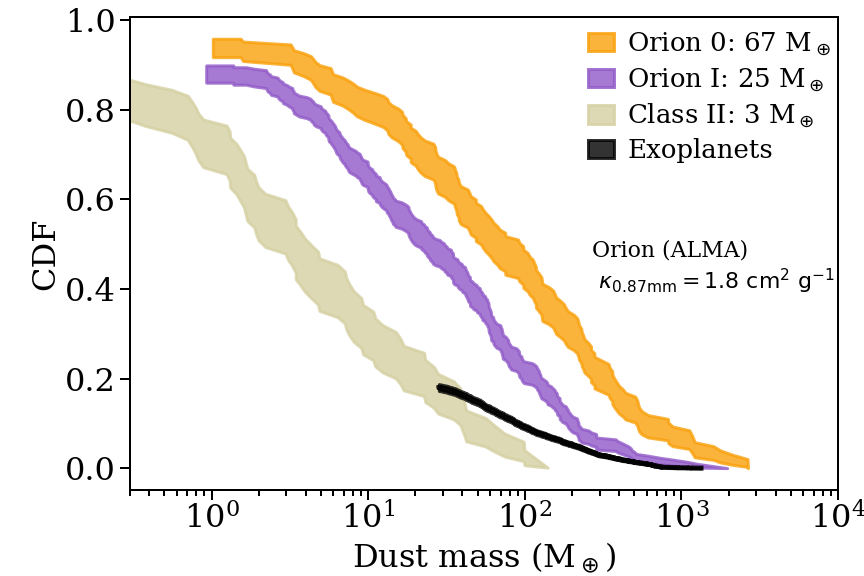}
    
  \caption{Cumulative distribution function of dust masses for Class 0 and Class I disks in Perseus and Orion and Class II disks in Lupus measured with ALMA {\it Top:} Perseus disk masses calculated with $\kappa_{1.3 {\rm mm}}=2.3\ {\rm cm^{2}\ g^{-1}}$ from the ALMA fluxes.  {\it Bottom: } Orion disk masses calculated with the $\kappa_{0.89 {\rm mm}}=1.3\ {\rm cm^{2}\ g^{-1}}$ \citep{Tobin2020}. Medians are indicated in the labels.}

 \label{fig:exo_efficiency}
\end{figure}

 \begin{figure*}[h]
\centering

    \includegraphics[width=0.95\linewidth,trim={0cm 0cm 0cm 0cm},clip]{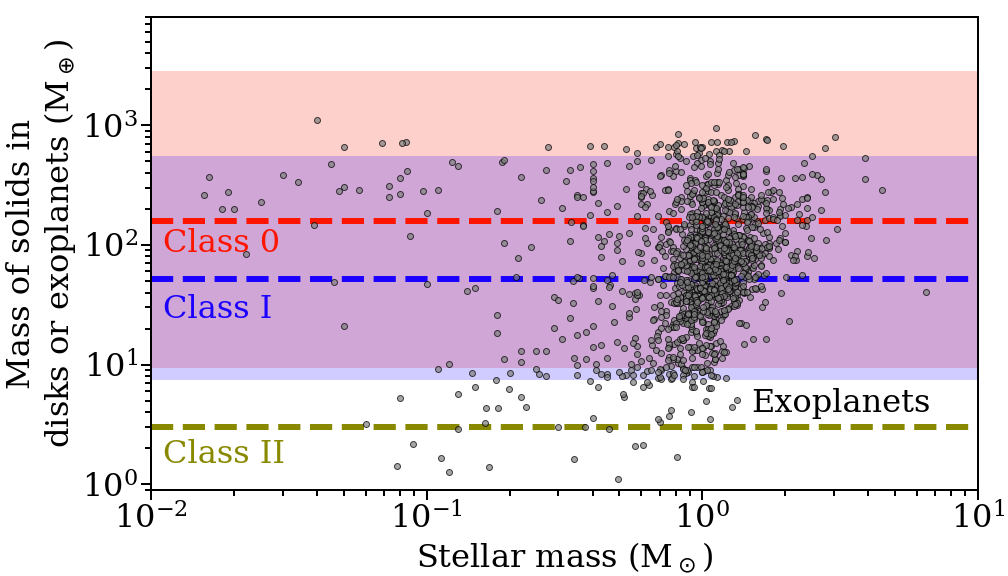}
  \caption{Plot showing the distribution of masses of exoplanetary systems obtained from the exoplanet.eu catalog \citep{Schneider2011}, for the planets around main-sequence stars with the measured masses. Shaded areas mark the range of our best estimation of the dust disk masses in Perseus: Class 0 (red) and Class I  (blue) calculated from the VLA fluxes with the opacity value of $\kappa_{9 {\rm mm}}=0.28\ {\rm cm^{2}\ g^{-1}}$. Medians of the distributions, 158 and 52 M$_\oplus$, for Class 0 and I, respectively, are indicated with the dashed lines. The median mass of the Class II disks in Lupus, 3  M$_\oplus$ \citep{Ansdell2016} is showed in yellow. The masses of the solids in exoplanetary systems are plotted against the stellar mass of the host star. All planets with available information on the mass are included in this plot, without introducing the 0.3 M$_{\rm J}$ threshold.}

 \label{fig:exo_scatter}
\end{figure*}

\subsection{Comparison of young disk dust masses with the solids in exoplanetary systems}

In Fig. \ref{fig:hist_exo+disks} (top panel) we compare the sample of exoplanetary systems with dust masses of young Class 0 and Class I disks in Perseus with our best estimate of embedded disk dust masses in Perseus, i.e., with the complete VLA survey, using the DIANA opacities. In Fig. \ref{fig:hist_exo+disks} (bottom panel) we show efficiency of planet formation for a given bin of the disk and exoplanet distributions. The efficiency is calculated as a ratio of the total mass of the exoplanetary systems at the certain fraction of the cumulative distribution plot divided by the corresponding dust mass of the disk at the same value of the CDF plot. This calculation provides information on how much total dust disk mass will be converted to planets.

In order to reproduce the population of exoplanets with the top 18\% most massive disks in Perseus, the efficiency of planet formation would have to be on the order of 15\% for Class 0 and 34\% for Class I (Fig.  \ref{fig:hist_exo+disks}). The average efficiency is measured by taking the mass at 10\% of the cumulative distribution plot (CDF plot is not well sampled for disk masses, and the 10\% value is the closest to the mean of the sample with data available). The Class 0/I disk masses in Perseus calculated from the VLA data at 9 mm for the refined value of the dust opacity suggest that on average there is enough mass available at those early phases to form the giant planet systems that we observe.

In Fig. \ref{fig:hist_exo+disks} (bottom panel), instead of the average value of the efficiency we attempt to measure the efficiency per each percentage level of the disk masses on the cumulative distribution plot . We note that this analysis has a higher uncertainty than the average value as the distribution of the exoplanets is uncertain. With known masses of a large number of giant planets and expected occurrences of such systems \citep{Cumming2008} we notice that the most massive exoplanets require efficiencies $\sim$ 30\%, stretching the requirements of some of the planet formation models (see Section 4.3). It is also possible that some of the most massive exoplanets or brown dwarfs present in the database do not follow the core accretion formation mechanisms, and excluding them would lower the requirement on efficiency.

If the underlying initial mass function of stars in Perseus is not representative of the stellar initial mass function, it could be that such exo-systems were produced from more massive disks than observed here. Another possibility is that young disks at the Class 0/I phase are still being replenished with material accreting from the envelope \citep{Hsieh2019}, making the effective material available for planet formation higher. It is also possible that the most massive systems are indeed producing the planets most efficiently.

In Fig. \ref{fig:exo_efficiency} (top panel) 
the disk masses from the ALMA fluxes in Perseus using the opacity value from \citet{Ansdell2016} are presented. We remind that this is likely the maximum value of the dust opacity at those wavelengths \citep{Panic2008} and therefore the masses are lower limits. In the bottom panel of Fig. \ref{fig:exo_efficiency}, the Orion Class 0/I disk masses measured with ALMA 0.87 mm observations \citep{Tobin2020} are used. 

The comparison of the exoplanet sample with the disk masses for the maximum value of the opacity used with the ALMA data (i.e., lower limit to the disk mass, Fig.  \ref{fig:exo_efficiency}, top panel) shows that if those opacities were the correct ones, the efficiency of forming planets in the Class 0 phase would be on the order of 33\%. The efficiency of the Class I phase would be on the order of 72\%. It is in line with our expectations that for our most conservative estimates of disk mass, the efficiencies required for giant planet formation are high, whereas for our best estimate of the dust masses in the young disks, we achieve an average efficiency in agreement with models (see Section 4.3). It is also clear that use of dust opacities that result in an order of magnitude lower disk masses \citep{Birnstiel2018} do not provide dust masses that would be compatible with such models.
The Orion sample, which contains more luminous protostars and could be more representative of the IMF than Perseus, has comparable disk dust masses in Class I and much lower disk masses in Class 0. The efficiencies required to produce the exoplanet population from the Orion disks dust content, as measured with ALMA, are comparable with those measured for Perseus with VLA observations (Fig. \ref{fig:exo_efficiency} bottom panel): 16\% and 45\% for Class 0 and Class I, respectively. 

Fig. \ref{fig:exo_scatter} presents a different visualization of the distribution of exoplanet systems dust masses compared to the range of disk masses observed in Class 0 and Class I disks in Perseus. The conclusions of our work show that for the complete sample of disks in Perseus and with a large sample of known exoplanets, there is enough solid material in Class 0 stage to explain the solid content in observed exoplanetary systems. This conclusion is consistent with \cite{Najita2014} and \cite{Greaves2011} but now with much more robust statistics.

In recent years several studies, specifically with the use of microlensing observations, estimate that nearly all stars can have at least 10 M{$_\oplus$} planet \citep{Cassan2012,Suzuki2016}. Very few cases and irreproducibility of microlensing observations suggest caution with extrapolating the results to all systems. It should be kept in mind that large population of wide-orbit planets could pose a challenge to efficiency of planet formation even in Class 0/I stage \citep{Najita2014}.

Actual timescales for the different phases of low-mass star formation are uncertain. Based on a statistical analysis of a population of protostars, it is estimated that the Class 0 evolutionary phase lasts for $\sim$0.1 Myr since the beginning of the collapse, and the Class I phase ends when the protostar is $\sim$ 0.5 Myr old \citep{Dunham2014}. More recent estimates of the half-lifes of the protostellar phases  give Class 0 half-life values of $\sim$ 0.05 Myr and $\sim$ 0.08 Myr for Class I \citep{Kristensen2018}. Therefore our results indicate the start of the planet formation begins less than 0.1 Myr after the beginning of the cloud collapse. This is consistent with the ages of the oldest meteorites in our Solar system \citep{Connelly2012}.

\subsection{The context of planet formation models}

It is of worth to put our empirical constraints on the solid mass reservoir in the context of planet formation models. Broadly speaking, planets can either form bottom-up through the assembly of smaller building blocks, in the so-called \textit{core accretion} scenario, or top-down in the \textit{gravitational instability} scenario, via direct gravitational collapse of the disk material. In the latter case (see \citealt{Kratter2016} for a review), planets need not contain rocky cores and we may therefore have over-estimated the solid mass locked in planets. In addition, in this view planets form at the very beginning of the disk lifetime when the disk is gravitationally unstable, possibly at even earlier stages than we probe here. If this is the case, our observations do not put constraints on the mass budget required for planet formation since planets would already be formed in the disks we are considering.

Our results are instead relevant for the core accretion scenario. In this case, two large families of models can be defined, differing in the type of building block: planetesimal accretion \citep[e.g.,][]{Pollack1996} and pebble accretion \citep{Ormel2010,Lambrechts2012}.

If planets grow by accreting planetesimals, it should be kept in mind that our observations are sensitive only to the dust. Therefore, the efficiency we have defined in this paper should be intended as the product of two efficiencies: the efficiency of converting dust into planetesimals and the efficiency of converting planetesimals into planets. The latter is relatively well constrained from theory and observations. Indeed, the accretion of planetesimals is highly efficient in numerical models \citep[e.g.,][]{Alibert2013} and nearly all planetesimals are accreted into planets over Myr timescales. Observationally, constraints on the mass in planetesimals that are not locked into planets is set by the debris disk population \citep{Sibthorpe2018}. For Sun-like stars, the median planetesimal mass\footnote{As discussed in \citet{Wyatt2007}, this value is degenerate with the maximum planetesimal size; here we have assumed a diameter of 1000 km, and we note that the mass would be even lower if using smaller planetesimals.} is 3 M$_\oplus$, i.e. much smaller than the solid content of giant planets. In this context, the Solar System could be an exception because attempts at explaining its complex history  (such as the well known Nice model, \citealt{Tsiganis2005}) require instead a much more massive (20-30 M$_\oplus$) population of planetesimals past the orbits of Uranus and Neptune that later evolved into the current Kuiper Belt. Even so, this mass is comparable to the mass in the cores of the giant planets, implying an efficiency of planet conversion from planetesimals $\gtrsim 50$\%.

Using this value, our observations place empirical constraints on the planetesimal formation efficiency. To satisfy our measurement of a total efficiency of $\sim 10$\%, a planetesimal formation efficiency of $\sim 20$\% would be needed. There is considerable uncertainty in planetesimal formation models \citep[e.g.,][]{Drazkowska2014, Drazkowska2018, Lenz2019}, but such an efficiency can in principle be reached by most of the expected conditions in the protoplanetary disks (see Fig. 9 of \citealt{Lenz2019} and Table 1 of \citealt{Drazkowska2014}). Thus, it is possible to explain the observed population of giant planets with the initial dust masses we report in this paper.

If instead planets grow by accreting pebbles, the growth rates can be significantly higher than in planetesimal accretion, but the formation efficiency is lower because most pebbles drift past the forming planets without being accreted \citep{Ormel2010,Ormel2017}. The efficiency of $\sim 10$\% reported here is among the highest that can be reached by pebble accretion \citep{Ormel2017} and it favours models where the disk is characterised by low turbulence ($\alpha < 10^{-3}$). 
Assuming such a value for the turbulence, \citet{Bitsch2019} finds that in pebble accretion models that form giant planets the total planet formation efficiency is 5-15\%, in line with our findings. It is also suggestive that in their models giant planet formation requires initial dust masses larger than $200-300$ M$_\oplus$: note how this condition is satisfied for $\sim$ 20\% of the Class 0 dust mass distribution. Therefore, the mass constraints derived in this paper from the VLA data are also consistent with pebble accretion, provided that the turbulence in the disk is sufficiently low.

\section{Conclusions}
This work collects available ALMA and VLA data of a complete sample of Perseus young disks in the Class 0/I phase to provide robust estimates of the disk dust masses at the early phases of star and planet formation. The refined values are used to compare the inferred disk masses with the exoplanetary systems to obtain constraints on when exoplanets start to form.

A linear correlation is found between the fluxes obtained with VLA and ALMA, supporting the fact that thermal dust emission is responsible for the emission at both wavelengths. The value of the dust spectral index measured  with ALMA and VLA observations is  $\beta  = 0.5 $, lower than the commonly used value of  $\beta  = 1 $, pointing to significant grain growth occurring already in the Class 0 and I phases. Therefore, compared with our previous study \citep{Tychoniec2018a} we recalculated the masses with the new dust opacity value that account for large grains. The best estimate of the median initial reservoir of dust mass available for planet formation in Perseus is 158 M$_\oplus$ and 52 M$_\oplus$ for Class 0 and Class I disks, respectively, derived from the VLA data. 

 Comparison of ALMA observations in Orion and Perseus shows that while disk masses in Class I disks agree well, Class 0 disks are more massive in Perseus than in Orion. This suggests that initial cloud conditions may lead to different masses of disks in the early phases.
 
Dust masses of disks measured with the VLA for Perseus are compared with the observed exoplanet systems.
If we assume that planet formation starts with the dust mass reservoir equal to the dust mass of Class I disks in Perseus, efficiency of $\sim$ 30\% is required to explain the currently observed systems with giant exoplanets. Lower efficiencies of $\sim$ 15\% on average are needed if the Class 0 disks are assumed as the starting point. We find strong evidence that there is enough dust mass in young disks to make planet formation possible already in the first $\sim$ 0.5 Myr of star formation. Given that low efficiencies are more in line with theoretical core accretion models, our results are most consistent with significant accumulation of material in larger bodies occurring already at the Class 0 phase.

\begin{acknowledgements}
{\L}T and EvD thank Dr. Yao Liu for discussions on dust opacities. {\L}T thanks Leon Trapman for discussions that helped in the presentation of results. G.R. acknowledges support from the Netherlands Organisation for Scientific Research (NWO, program number 016.Veni.192.233). JT acknowledges support from grant AST-1814762 from the National Science Foundation. The National Radio Astronomy Observatory is a facility of the National Science Foundation operated under cooperative agreement by Associated Universities, Inc. This paper makes use of the following ALMA data: ADS/JAO.ALMA\#2017.1.01693.S,  ADS/JAO.ALMA\#2017.1.01078.S, ADS/JAO.ALMA\#2015.1.00041.S, and ADS/JAO.ALMA\#2013.1.00031.S. ALMA is a partnership of ESO (representing its member states), NSF (USA) and NINS (Japan), together with NRC (Canada), MOST and ASIAA (Taiwan), and KASI (Republic of Korea), in cooperation with the Republic of Chile. The Joint ALMA Observatory is operated by ESO, AUI/NRAO and NAOJ. Astrochemistry in Leiden is supported by the Netherlands Research School for Astronomy (NOVA), by a Royal Netherlands Academy of Arts and Sciences (KNAW) professor prize, and by the European Union A-ERC grant 291141 CHEMPLAN. This project has received funding from the European Union's Horizon 2020 research and innovation programme under the Marie Skłodowska-Curie grant agreement No 823823 (DUSTBUSTERS). This work was partly supported by the Deutsche Forschungs-Gemeinschaft (DFG, German Research Foundation) - Ref no. FOR 2634/1 TE 1024/1-1. This research made use of Astropy, a community-developed core Python package for Astronomy \citep{Astropy2013},
http://astropy.org); Matplotlib library \citep{Hunter2007}; NASA's Astrophysics Data System.
\end{acknowledgements}

\bibliography{mybib}


\longtab{
\begin{longtable}{lrrrrrrrr}
\caption{Measured properties of the disks in Perseus molecular cloud}\\
\hline

\label{tab:table1}

Source name & $F_{\lambda}$& Size & Deconvolved size & Envelope & Residual  & Ext. fit \\
  & mJy & $\arcsec$ & $\arcsec$ &   \%  & \% &\\
\endfirsthead
\hline \hline
Per-emb-1 & $57.5\pm0.6$ & $0.48\times0.34$ & $0.27\times0.16$ &  11 &  <1  & yes \\
Per-emb-2 & $593.4\pm13.3$ & $1.24\times0.86$ & $1.16\times0.79$ &  20 &  <1  & yes \\
Per-emb-3 & $53.0\pm0.3$ & $0.53\times0.36$ & $0.27\times0.16$ &  4 &  <1  & yes \\
Per-emb-4 & $0.5\pm0.1$ & \tablefoottext{a}$0.47\times0.31$ & --- & <1 &  <1 & no \\
Per-emb-5 & $276.8\pm1.5$ & $0.59\times0.40$ & $0.37\times0.25$ &  3 &  <1  & yes \\
Per-emb-6 & $11.4\pm0.2$ & $0.50\times0.34$ & $0.20\times0.13$ &  4 &  <1  & yes \\
Per-emb-7 & $2.5\pm0.3$ & $0.83\times0.59$ & $0.75\times0.41$ &  47 &  <1  & yes \\
Per-emb-9 & $10.0\pm0.3$ & $0.66\times0.45$ & $0.48\times0.32$ &  27 &  <1  & yes \\
Per-emb-10 & $22.5\pm0.2$ & $0.50\times0.34$ & $0.18\times0.12$ &  4 &  <1  & yes \\
Per-emb-11-A & $160.3\pm0.8$ & $0.52\times0.42$ & $0.32\times0.30$ &  4 &  <1  & yes \\
Per-emb-11-B & $5.1\pm0.5$ & $0.41\times0.28$ & --- & <1 &  <1 & no \\
Per-emb-11-C & $3.5\pm0.2$ & $0.48\times0.35$ & $0.25\times0.21$ &  26 &  <1  & yes \\
Per-emb-14 & $98.8\pm0.6$ & $0.62\times0.36$ & $0.42\times0.17$ &  3 &  <1  & yes \\
Per-emb-15 & $6.5\pm0.3$ & $0.54\times0.42$ & $0.29\times0.27$ &  19 &  <1  & yes \\
Per-emb-19 & $17.7\pm0.2$ & $0.50\times0.34$ & $0.18\times0.14$ &  2 &  <1  & yes \\
Per-emb-20 & $8.1\pm0.3$ & $0.68\times0.50$ & $0.50\times0.38$ &  29 &  <1  & yes \\
Per-emb-22-A & $49.0\pm1.1$ & $0.66\times0.43$ & $0.49\times0.27$ &  28 &  1  & yes \\
Per-emb-22-B & $24.7\pm1.1$ & $0.66\times0.41$ & $0.49\times0.23$ &  \tablefoottext{b}{60} &  \tablefoottext{b}2  & yes \\
Per-emb-24 & $3.9\pm0.3$ & $0.60\times0.40$ & $0.39\times0.24$ &  1 &  1  & no \\
Per-emb-25 & $126.6\pm0.8$ & $0.57\times0.36$ & $0.33\times0.18$ &  <1 &  <1 & no \\
Per-emb-27-A & $222.4\pm3.7$ & $0.47\times0.44$ & --- &  13 &  1  & yes \\
Per-emb-27-B & $23.3\pm3.5$ & $0.44\times0.44$ & --- &  \tablefoottext{b}{224} &  \tablefoottext{b}{56}  & yes \\
Per-emb-29 & $127.6\pm1.7$ & $0.52\times0.37$ & $0.23\times0.19$ &  11 &  <1  & yes \\
Per-emb-30 & $48.4\pm0.3$ & $0.49\times0.35$ & $0.18\times0.13$ &  3 &  <1  & yes \\
Per-emb-31 & $1.1\pm0.2$ & $0.43\times0.34$ & --- &  17 &  <1  & yes \\
Per-emb-34 & $12.4\pm0.2$ & $0.52\times0.36$ & $0.24\times0.18$ &  6 &  <1  & yes \\
Per-emb-35-B & $19.9\pm0.3$ & $0.50\times0.33$ & $0.19\times0.11$ &  11 &  <1  & yes \\
Per-emb-35-A & $27.8\pm0.3$ & $0.50\times0.34$ & $0.20\times0.14$ &  7 &  <1  & yes \\
Per-emb-38 & $27.7\pm0.2$ & $0.52\times0.38$ & $0.27\times0.16$ &  <1 &  <1 & no \\
Per-emb-39 & $0.9\pm0.2$ & $0.67\times0.34$ & $0.48\times0.09$ &  20 &  1  & yes \\
Per-emb-41 & $11.3\pm0.2$ & $0.47\times0.34$ & $0.11\times0.10$ &  <1 &  <1 & no \\
Per-emb-45 & $1.0\pm0.1$ & $0.45\times0.34$ & --- &  9 &  <1  & yes \\
Per-emb-46 & $6.0\pm0.2$ & $0.58\times0.38$ & $0.35\times0.22$ &  2 &  1  & no \\
Per-emb-49-A & $22.1\pm0.2$ & $0.49\times0.35$ & $0.18\times0.13$ &  2 &  <1  & yes \\
Per-emb-49-B & $4.7\pm0.2$ & $0.50\times0.30$ & --- &  46 &  <1  & yes \\
Per-emb-50 & $86.4\pm0.2$ & $0.46\times0.29$ & $0.22\times0.07$ &  <1 &  <1 & no \\
Per-emb-51 & $1.7\pm0.6$ & $0.48\times0.43$ & $0.29\times0.10$ &  262 &  5  & yes \\
Per-emb-52 & $5.2\pm0.2$ & $0.52\times0.34$ & $0.23\times0.13$ &  <1 &  <1 & no \\
Per-emb-53 & $29.5\pm0.4$ & $0.54\times0.42$ & $0.39\times0.34$ &  <1 &  <1 & no \\
Per-emb-54 & $3.1\pm0.6$ & $0.59\times0.50$ & $0.39\times0.36$ &  <1 &  <1 & no \\
Per-emb-58 & $4.6\pm0.2$ & $0.51\times0.36$ & $0.22\times0.16$ &  <1 &  <1 & no \\
Per-emb-59 & $1.6\pm0.1$ & $0.47\times0.32$ & --- & <1 &  <1 & no \\
Per-emb-62 & $77.0\pm0.1$ & $0.44\times0.33$ & $0.19\times0.13$ &  <1 &  <1 & no \\
Per-emb-63-B & $3.8\pm0.2$ & $0.53\times0.35$ & $0.26\times0.16$ &  <1 &  <1 & no \\
Per-emb-63-C & $2.2\pm0.2$ & $0.50\times0.31$ & --- & <1 &  <1 & no \\
Per-emb-63-A & $25.8\pm0.2$ & $0.50\times0.36$ & $0.18\times0.17$ &  <1 &  <1 & no \\
Per-emb-64 & $42.4\pm0.3$ & $0.51\times0.34$ & $0.20\times0.13$ &  <1 &  <1 & no \\
Per-emb-65 & $37.1\pm0.2$ & $0.53\times0.39$ & $0.26\times0.23$ &  <1 &  <1 & no \\
B1-bS & $330.4\pm2.6$ & $0.60\times0.52$ & $0.42\times0.39$ &  7 &  <1  & yes \\
SVS13C & $62.1\pm0.9$ & $0.46\times0.38$ & $0.29\times0.09$ &  <1 &  <1 & no \\
SVS13B & $207.3\pm5.9$ & $0.57\times0.47$ & $0.37\times0.31$ &  20 &  <1  & yes \\
SVS13A2 & $18.8\pm3.8$ & \tablefoottext{a}$0.46\times0.31$ & --- & <1 &  <1 & no \\
\hline
\end{longtable}
\tablefoot{
$F_\lambda$- integrated flux density at the wavelength of the observations: 1.1 mm for sources in the \cite{Hsieh2019}, 1.3 mm for sources in the 2017.1.01078.S dataset (PI: D. Segura-Cox), uncertainty is provided by the {\it imfit} task in CASA; Size - FWHM of the 2D Gaussian fit to the disk component; Deconvolved size - the size of the emission deconvolved from the beam, provided by {\it imfit} task in CASA, where no value is provided, the size of the emission is too close to the synthesized beam size; Envelope - flux remaining in the ellipse of the size of the disk, after removing the model of the disk from the image as a fraction of the disk flux; Residual -  flux remaining in the ellipse of the size of the disk, after removing both the disk model and the broad envelope model, as a fraction of the disk flux; Ext. fit - flag indicating if the second, broad Gaussian component was used in the fit.
\tablefoottext{a}{FWHM of the Gaussian fixed before fitting.}
\tablefoottext{b}{Value not reliable due to a close companion.}
}
}

\longtab{
\begin{longtable}{lcrrrrrr}
\caption{Masses of the embedded disks in Perseus}\\
\hline

Source name & Class  & $M_{\rm dust}$@ 9 mm & $M_{\rm dust}$ @ 1 mm & $\alpha$ & $\alpha$ & robust & Dataset \\
 & & M$_\oplus$ &   M$_\oplus$ & VLA/ALMA & VLA \\\hline
  \hline
\endfirsthead
\caption{continued.}\\
\hline
\label{tab:table2}

Source name & Class  & $M_{\rm dust}$@ 9 mm & $M_{\rm dust}$ @ 1 mm & $\alpha$ & $\alpha$ & robust & Dataset \\
 & & M$_\oplus$ &   M$_\oplus$ & VLA/ALMA & VLA \\\hline
 \hline
 \hline
\endhead
\endfoot
\hline
Per-emb-1 & 0  &  $279.6\pm50.0$ &  $82.5\pm1.2$ & 2.4 & 1.9 & n & Segura-Cox   \\
Per-emb-2 & 0  &  $927.4\pm175.6$ &  $573.6\pm18.2$ & 2.8 & 2.9 & y & Hsieh   \\
Per-emb-3 & 0  &  $157.8\pm32.7$ &  $51.2\pm0.4$ & 2.4 & 1.5 & n & Hsieh   \\
Per-emb-4 & 0  & <$11.3\pm7.2$ &  $0.5\pm0.1$ & 1.5 &---& n & Hsieh   \\
Per-emb-5 & 0  &  $502.3\pm86.3$ &  $267.6\pm2.1$ & 2.7 & 2.7 & y & Hsieh   \\
Per-emb-6 & 0  &  $78.5\pm21.5$ &  $11.0\pm0.2$ & 2.0 & 1.7 & y & Hsieh   \\
Per-emb-7 & 0  & <$11.8\pm7.2$ &  $2.4\pm0.4$ & 2.2 &---& n & Hsieh   \\
Per-emb-8 & 0  &  $237.9\pm47.5$ &  $147.6\pm5.5$ & 2.7 & 3.0 & y & Tobin   \\
Per-emb-9 & 0  &  $23.1\pm14.4$ &  $9.7\pm0.4$ & 2.6 & 3.3 & n & Hsieh   \\
Per-emb-10 & 0  &  $143.4\pm30.7$ &  $21.7\pm0.3$ & 2.1 & 1.5 & n & Hsieh   \\
Per-emb-11-A & 0  &  $413.5\pm73.1$ &  $230.2\pm1.6$ & 2.7 & 2.4 & y & Segura-Cox   \\
Per-emb-11-B & 0  &  $21.7\pm14.9$ &  $7.4\pm1.1$ & 2.4 & 1.7 & n & Segura-Cox   \\
Per-emb-11-C & 0  &  $30.5\pm15.4$ &  $5.1\pm0.4$ & 2.1 & 3.2 & n & Segura-Cox   \\
Per-emb-12-A & 0  &  $2853.9\pm437.0$ &  $2159.5\pm59.3$ & 2.8 & 3.0 & y & Tobin   \\
Per-emb-12-B & 0  &  $158.1\pm38.4$ &  $623.6\pm16.0$ & 3.7 & 3.2 & n & Tobin   \\
Per-emb-13 & 0 &  $1271.3\pm207.6$ & ---   & --- & 2.7 & n & --- \\
Per-emb-14 & 0  &  $311.2\pm58.6$ &  $95.5\pm0.9$ & 2.4 & 2.4 & y & Hsieh   \\
Per-emb-15 & 0  & <$9.3\pm6.6$ &  $6.3\pm0.4$ & 2.8 &---& n & Hsieh   \\
Per-emb-16 & 0  & <$34.9\pm18.1$  & --- & --- &  --- &  n &  ---   \\
Per-emb-17-A & 0  &  $160.1\pm32.4$ &  $37.5\pm1.0$ & 2.2 & 1.9 & y & Tobin   \\
Per-emb-17-B & 0  &  $39.3\pm10.4$ &  $29.3\pm1.0$ & 2.8 & 2.6 & y & Tobin   \\
Per-emb-18 & 0  &  $224.8\pm47.1$ &  $208.5\pm2.8$ & 2.9 & 1.5 & n & Tobin   \\
Per-emb-19 & 0  &  $112.3\pm27.7$ &  $17.1\pm0.2$ & 2.1 & 1.1 & n & Hsieh   \\
Per-emb-20 & 0  & <$31.1\pm15.5$ &  $7.8\pm0.4$ & 2.3 &---& n & Hsieh   \\
Per-emb-21 & 0  &  $211.9\pm41.1$ &  $70.8\pm1.0$ & 2.4 & 2.5 & y & Tobin   \\
Per-emb-22-A & 0  &  $96.4\pm24.1$ &  $47.3\pm1.5$ & 2.6 & 1.0 & n & Hsieh   \\
Per-emb-22-B & 0  &  $23.1\pm13.0$ &  $23.9\pm1.5$ & 3.0 & 3.4 & n & Hsieh   \\
Per-emb-23 & 0 &  $37.2\pm15.0$ & ---   & --- & 3.8 & n & --- \\
Per-emb-24 & 0  &  $21.5\pm7.9$ &  $3.8\pm0.3$ & 2.2 & 3.0 & n & Hsieh   \\
Per-emb-25 & 0  &  $172.6\pm65.4$ &  $122.4\pm1.1$ & 2.8 & 1.7 & n & Hsieh   \\
Per-emb-26 & 0 &  $636.1\pm102.5$ & ---   & --- & 2.2 & n & --- \\
Per-emb-27-A & 0  &  $570.1\pm90.4$ &  $215.0\pm5.1$ & 2.5 & 1.9 & n & Hsieh   \\
Per-emb-27-B & 0  &  $113.6\pm24.4$ &  $22.5\pm4.8$ & 2.2 & 1.9 & y & Hsieh   \\
Per-emb-28 & 0/I &  $23.5\pm16.6$ & ---   & --- & 4.4 & n & --- \\
Per-emb-29 & 0/I  &  $233.5\pm43.5$ &  $123.3\pm2.4$ & 2.7 & 1.8 & n & Hsieh   \\
Per-emb-30 & 0/I  &  $263.5\pm47.7$ &  $46.8\pm0.4$ & 2.2 & 2.1 & y & Hsieh   \\
Per-emb-31 & 0/I  & <$26.5\pm14.9$ &  $1.1\pm0.2$ & 1.5 &---& n & Hsieh   \\
Per-emb-32-A & 0/I &  $18.2\pm8.1$ & ---   & --- & 0.9 & n & --- \\
Per-emb-32-B & 0/I &  $30.0\pm10.5$ & ---   & --- & 3.6 & n & --- \\
Per-emb-33-A & 0  &  $294.0\pm57.6$ &  $9.3\pm1.6$ & 1.2 & 2.3 & n & Tobin   \\
Per-emb-33-B & 0  &  $170.6\pm39.4$ &  $45.8\pm2.8$ & 2.3 & 1.2 & n & Tobin   \\
Per-emb-33-C & 0  &  $55.8\pm19.0$ &  $234.1\pm3.7$ & 3.7 & 2.0 & n & Tobin   \\
Per-emb-34 & I  &  $85.6\pm22.5$ &  $12.0\pm0.3$ & 2.0 & 1.2 & n & Hsieh   \\
Per-emb-35-B & I  &  $86.8\pm20.6$ &  $19.3\pm0.4$ & 2.3 & 1.3 & n & Hsieh   \\
Per-emb-35-A & I  &  $47.0\pm15.7$ &  $26.8\pm0.4$ & 2.7 & 3.2 & n & Hsieh   \\
Per-emb-36-A & I  &  $555.9\pm90.4$ &  $185.5\pm1.8$ & 2.4 & 2.5 & y & Tobin   \\
Per-emb-36-B & I  &  $110.6\pm23.3$ &  $17.7\pm1.4$ & 2.0 & 2.2 & y & Tobin   \\
Per-emb-37 & 0 &  $95.4\pm23.9$ & ---   & --- & 1.7 & n & --- \\
Per-emb-38 & I  &  $78.7\pm23.6$ &  $26.7\pm0.3$ & 2.5 & 1.9 & n & Hsieh   \\
Per-emb-39 & I  & <$12.3\pm7.3$ &  $0.8\pm0.2$ & 1.7 &---& n & Hsieh   \\
Per-emb-40-A & I  &  $72.9\pm18.0$ &  $22.8\pm0.6$ & 2.4 & 2.3 & y & Tobin   \\
Per-emb-40-B & I  & <$17.6\pm8.7$ &  $1.3\pm0.2$ & 1.7 &---& n & Tobin   \\
Per-emb-41 & I  &  $44.2\pm18.8$ &  $11.0\pm0.3$ & 2.3 & 0.9 & n & Hsieh   \\
Per-emb-42 & I &  $116.5\pm29.0$ & ---   & --- & 1.7 & n & --- \\
Per-emb-43 & I  & <$11.8\pm7.2$  & --- & --- &  --- &  n &  ---   \\
Per-emb-44-A & 0/I  &  $497.7\pm80.8$ &  $142.2\pm2.2$ & 2.3 & 2.3 & y & Tobin   \\
Per-emb-44-B & 0/I  &  $176.7\pm41.7$ &  $227.1\pm3.7$ & 3.1 & 1.7 & n & Tobin   \\
Per-emb-45 & I  & <$11.8\pm7.2$ &  $1.0\pm0.2$ & 1.8 &---& n & Hsieh   \\
Per-emb-46 & I  &  $55.5\pm26.6$ &  $5.8\pm0.3$ & 1.9 & 0.8 & n & Hsieh   \\
Per-emb-47 & I &  $106.5\pm28.3$ & ---   & --- & 2.2 & n & --- \\
Per-emb-48-A & I  &  $37.9\pm16.5$ &  $5.7\pm0.8$ & 2.0 & 0.7 & n & Tobin   \\
Per-emb-48-B & I  & <$14.9\pm10.5$ & <$0.6\pm0.8$  & 1.3 & --- & n &  Tobin   \\
Per-emb-49-A & I  &  $145.2\pm31.0$ &  $21.4\pm0.3$ & 2.1 & 0.6 & n & Hsieh   \\
Per-emb-49-B & I  & <$53.0\pm16.5$ &  $4.5\pm0.3$ & 1.8 &---& n & Hsieh   \\
Per-emb-50 & I  &  $535.2\pm91.0$ &  $124.1\pm0.4$ & 2.2 & 2.1 & y & Segura-Cox   \\
Per-emb-51 & I  & <$12.8\pm8.1$ &  $1.6\pm0.8$ & 2.0 &---& n & Hsieh   \\
Per-emb-52 & I  &  $35.4\pm19.6$ &  $5.0\pm0.3$ & 2.0 & 1.0 & n & Hsieh   \\
Per-emb-53 & I  &  $56.6\pm22.6$ &  $42.3\pm0.8$ & 2.8 & 1.1 & n & Segura-Cox   \\
Per-emb-54 & I  &  $115.5\pm28.2$ &  $3.0\pm0.9$ & 1.2 & 1.0 & y & Hsieh   \\
Per-emb-55-A & I  &  $48.7\pm11.8$ &  $4.7\pm1.0$ & 1.8 & 3.6 & n & Tobin   \\
Per-emb-55-B & I  &  $7.4\pm5.2$ &  $0.7\pm0.6$ & 1.8 & 4.4 & n & Tobin   \\
Per-emb-56 & I &  $45.7\pm16.9$ & ---   & --- & 3.7 & n & --- \\
Per-emb-57 & I &  $67.9\pm22.8$ & ---   & --- & 0.6 & n & --- \\
Per-emb-58 & I  & <$37.9\pm17.9$ &  $4.4\pm0.2$ & 2.0 &---& n & Hsieh   \\
Per-emb-59 & I  & <$13.8\pm8.2$ &  $1.5\pm0.2$ & 1.9 &---& n & Hsieh   \\
Per-emb-60 & I  & <$13.3\pm8.1$  & --- & --- &  --- &  n &  ---   \\
Per-emb-61 & I &  $52.1\pm23.4$ & ---   & --- & 2.6 & n & --- \\
Per-emb-62 & I  &  $200.8\pm41.6$ &  $110.5\pm0.2$ & 2.7 & 2.2 & n & Segura-Cox   \\
Per-emb-63-B & I  & <$11.8\pm8.3$ &  $3.6\pm0.3$ & 2.4 &---& n & Hsieh   \\
Per-emb-63-C & I  & <$11.8\pm8.3$ &  $2.1\pm0.2$ & 2.2 &---& n & Hsieh   \\
Per-emb-63-A & I  &  $109.7\pm27.3$ &  $24.9\pm0.3$ & 2.3 & 2.0 & y & Hsieh   \\
Per-emb-64 & I  &  $240.9\pm48.0$ &  $40.9\pm0.4$ & 2.1 & 1.8 & y & Hsieh   \\
Per-emb-65 & I  &  $58.5\pm25.0$ &  $35.9\pm0.3$ & 2.8 & 1.1 & n & Hsieh   \\
Per-emb-66 & I  & <$13.3\pm8.1$  & --- & --- &  --- &  n &  ---   \\
Per-bolo-58 & 0  & <$12.8\pm8.1$  & --- & --- &  --- &  n &  ---   \\
Per-bolo-45 & 0  & <$12.3\pm7.3$  & --- & --- &  --- &  n &  ---   \\
L1451-MMS & 0 &  $97.3\pm22.8$ & ---   & --- & 2.2 & n & --- \\
L1448IRS2E & 0  & <$14.7\pm9.0$  & --- & --- &  --- &  n &  ---   \\
B1-bN & 0 &  $483.8\pm85.1$ & ---   & --- & 2.5 & n & --- \\
B1-bS & 0  &  $354.0\pm76.5$ &  $319.4\pm3.5$ & 2.9 & 2.1 & n & Hsieh   \\
L1448IRS1-A & I  &  $292.7\pm53.2$ &  $105.4\pm1.8$ & 2.5 & 2.2 & y & Tobin   \\
L1448IRS1-B & I  &  $36.4\pm15.6$ &  $7.9\pm1.4$ & 2.2 & 3.7 & n & Tobin   \\
L1448NW-A & 0  &  $194.5\pm38.6$ &  $71.5\pm1.0$ & 2.5 & 1.7 & n & Tobin   \\
L1448NW-B & 0  &  $187.9\pm39.1$ &  $21.3\pm0.4$ & 1.9 & 1.7 & y & Tobin   \\
L1448IRS3A & I  &  $235.1\pm47.8$ &  $145.0\pm5.1$ & 2.7 & 0.7 & n & Tobin   \\
SVS13C & 0  &  $223.9\pm48.4$ &  $89.3\pm1.9$ & 2.5 & 1.4 & n & Segura-Cox   \\
SVS13B & 0  &  $581.1\pm98.2$ &  $200.3\pm8.1$ & 2.5 & 2.3 & y & Hsieh   \\
IRAS03363+3207 & I? &  $326.3\pm58.0$ & ---   & --- & 2.4 & n & --- \\
IRAS4B' & 0 &  $603.2\pm115.2$ & ---   & --- & 2.6 & n & --- \\
SVS13A2 & 0?  &  $102.6\pm27.7$ &  $18.2\pm5.2$ & 2.2 & 1.4 & n & Hsieh   \\
\hline
\end{longtable}
\tablefoot{
Class - Evolutionary class of the source. In the analysis, borderline sources (0/I) are treated as Class 0. Dataset: Tobin -  \citep{Tobin2018} (1.3 mm), Segura-Cox - 2017.1.01078.S (1.3 mm), Hsieh - \citep{Hsieh2019} (1.1 mm).  $\alpha_{\rm VLA/ALMA}$  - spectral indices obtained  between 1 and 9 mm, $\alpha_{\rm VLA}$ - Ka-band intraband indices. Robust - if 'y', sources with consistent spectral indices betweem VLA/ALMA and VLA Ka-band.} 
}

\appendix
\section{Additional plots}

\begin{figure*}[h]
\centering
     \includegraphics[width=0.9\linewidth,trim={5cm 0cm 3cm 0cm},clip]{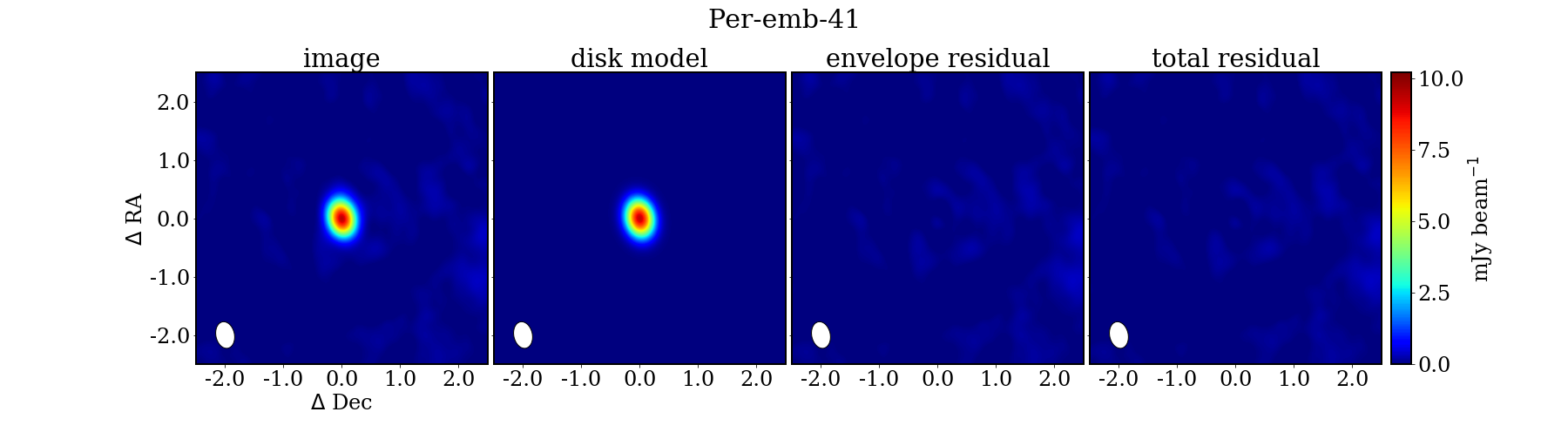}
     \includegraphics[width=0.9\linewidth,trim={5cm 0cm 3cm 0cm},clip]{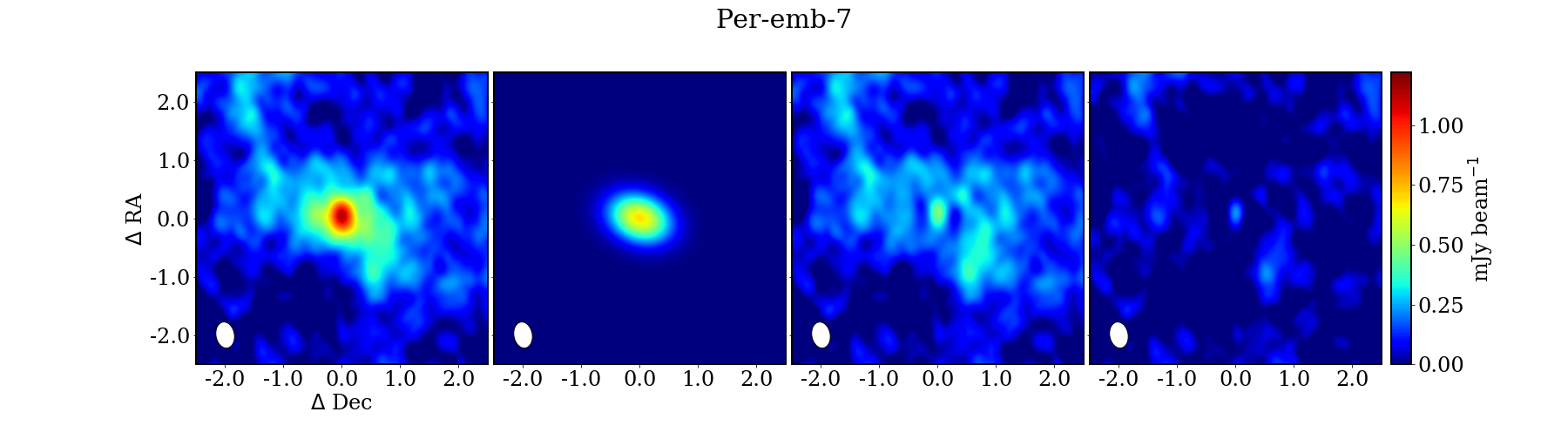}
    \includegraphics[width=0.9\linewidth,trim={5cm 0cm 3cm 0cm},clip]{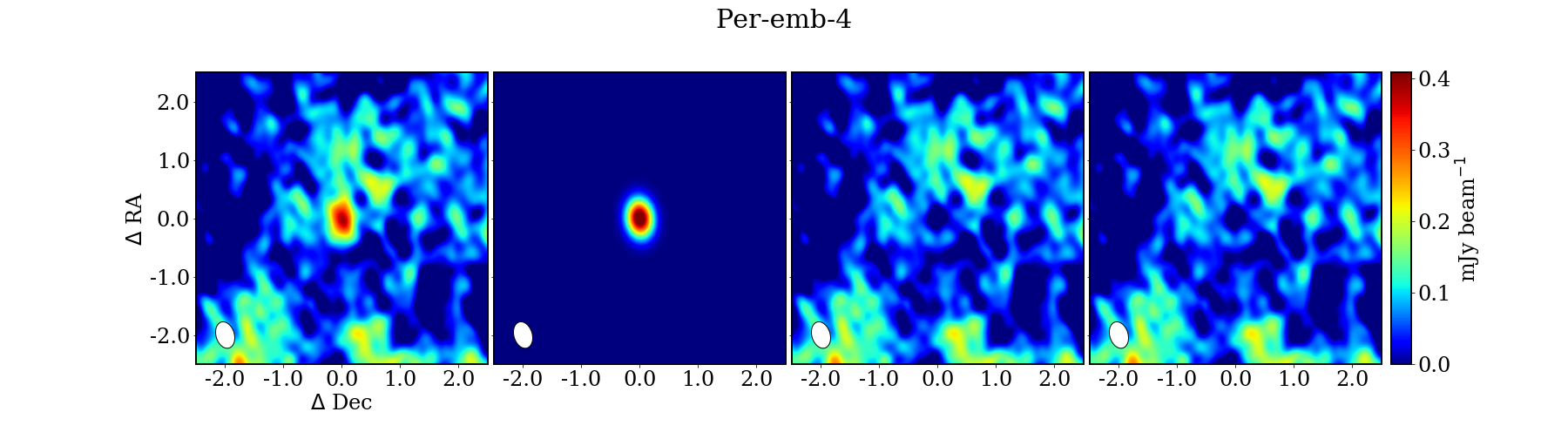}
     \includegraphics[width=0.9\linewidth,trim={5cm 0cm 3cm 0cm},clip]{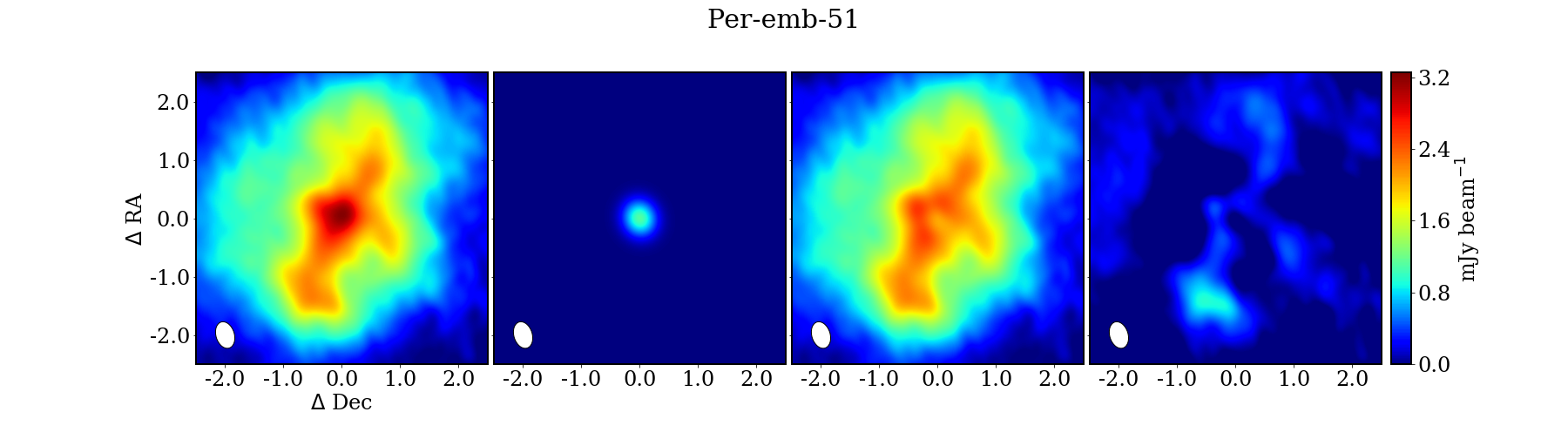}

  \caption{From left to right: observed 1.1 mm ALMA continuum images of the targeted protostars; an image of the disk model for the source resulting from the Gaussian fitting, in cases where the envelope component is fitted, only the disk component is shown; the residual image after subtracting the disk model from the image; the residual image after subtracting both disk and envelope models. From top to bottom, cases that occur in the Gaussian fitting procedure are presented:  Per-emb-41, single source, without a noticeable contribution from the envelope; Per-emb-7, contribution of the envelope, as assessed by eye, is significant, an additional broad Gaussian with size of 3\arcsec was added to the input parameters; Per-emb-54, the central compact component was faint compared to the noise level, it was necessary to fix the size of the Gaussian to the synthesized beam size for the fit to converge. Per-emb-51, the central component has a very small flux compared to the envelope. }
 \label{fig:fit_examples}
\end{figure*}

\begin{figure*}[h]
\centering
  \includegraphics[width=0.33\linewidth,trim={0 0cm 0 0cm},clip]{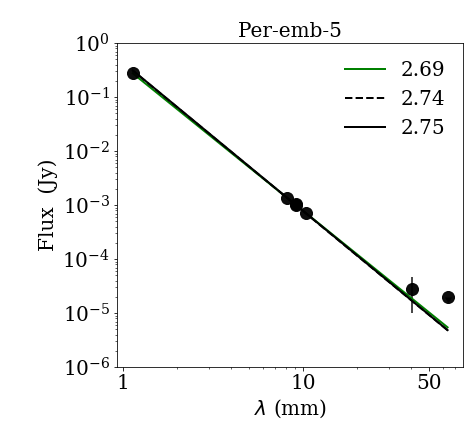}
      \includegraphics[width=0.33\linewidth,trim={0 0cm 0 0cm},clip]{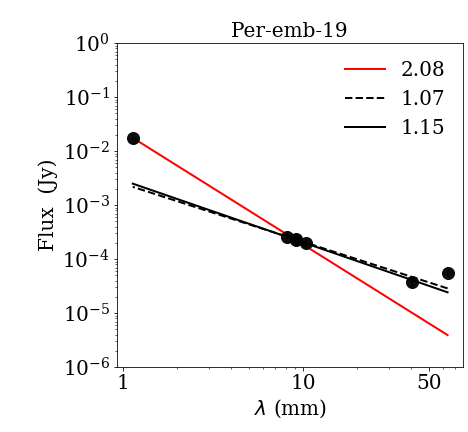}
    \includegraphics[width=0.33\linewidth,trim={0 0cm 0 0cm},clip]{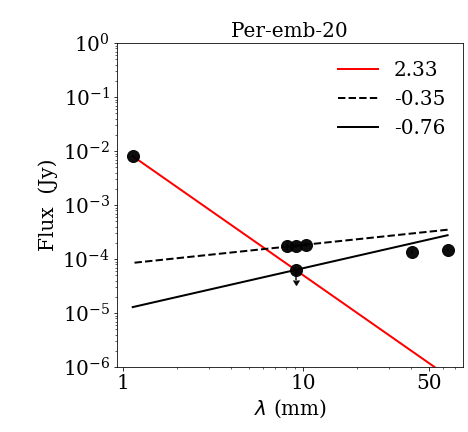}
  \caption{Examples of fitting the spectral index to the fluxes from VLA (8 mm - 6.4 cm) and ALMA observations (1 mm). {\it Left:} Per-emb-5 has an intra-band Ka-band spectral index in agreement with the spectral index between 1 mm and 9 mm wavelengths. {\it Center:} For Per-emb-19 Ka-band spectral index is too low to be explained only by dust emission, some free-free contamination cannot be excluded. {\it Right:} In Per-emb-20, negative spectral index at Ka-band shows that emission is heavily contaminated by other effects, in such cases the Ka-band flux is treated as an upper limit of the dust emission.}
 \label{fig:SED}
\end{figure*}

\begin{figure*}[h]
\centering
  \includegraphics[width=0.45\linewidth,trim={0 0cm 0 0cm},clip]{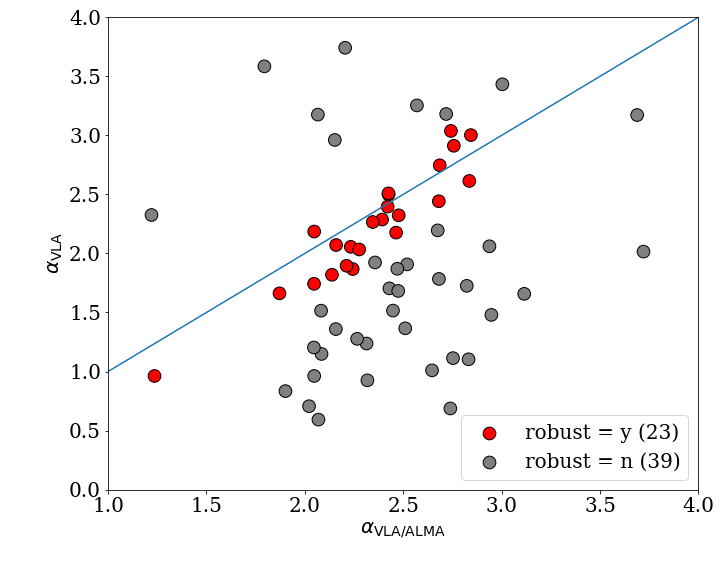}

  \caption{Plot showing spectral indices obtained  between 1 and 9 mm $\alpha_{\rm VLA/ALMA}$ and Ka-band intraband indices $\alpha_{\rm VLA}$, only for sources with both values provided. Red points are the sources with consistent spectral indices betweem VLA/ALMA and VLA Ka-band.}
 \label{fig:robust_index}
\end{figure*}

\begin{figure*}[h]
\centering
  \includegraphics[width=0.45\linewidth,trim={0 0cm 0 0cm},clip]{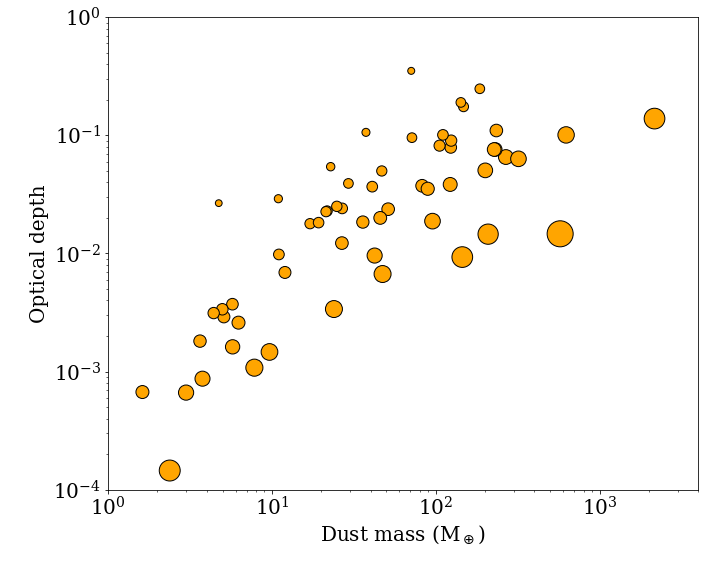}

  \caption{Plot showing optical depth as a function of the disk dust masses. Size of the circle is proportional to the disk size of the source.}
 \label{fig:dustopacities}
\end{figure*}

 \end{document}